# Fast ATP-dependent Subunit Rotation in Reconstituted $F_oF_1$-ATP Synthase Trapped in Solution

*Thomas Heitkamp, Michael Börsch\**

Single-Molecule Microscopy Group, Jena University Hospital, 07743 Jena, Germany.

KEYWORDS: single-molecule enzymology, $F_oF_1$-ATP synthase, subunit rotation, single-molecule FRET, ABEL trap, ATP hydrolysis, proton motive force.

\* corresponding author: michael.boersch@med.uni-jena.de

ABSTRACT: $F_oF_1$-ATP synthases are ubiquitous membrane-bound, rotary motor enzymes that can catalyze ATP synthesis and hydrolysis. Their enzyme kinetics are controlled by internal subunit rotation, by substrate and product concentrations, by mechanical inhibitory mechanisms, but also by the electrochemical potential of protons across the membrane. Single-molecule Förster resonance energy transfer (smFRET) has been used to detect subunit rotation within $F_oF_1$-ATP synthases embedded in freely diffusing liposomes. We now report that kinetic monitoring of functional rotation can be prolonged from milliseconds to seconds by utilizing an Anti-Brownian electrokinetic trap (ABEL trap). These extended observation times allowed us to observe fluctuating rates of functional rotation for individual $F_oF_1$-liposomes in solution. Broad distributions of ATP-dependent catalytic rates were revealed. The buildup of an electrochemical potential of protons was confirmed to limit the maximum rate of ATP hydrolysis. In the presence of ionophores or uncouplers, the fastest subunit rotation speeds measured in single reconstituted $F_oF_1$-ATP synthases were 180 full rounds per second. This was much faster than measured by biochemical ensemble averaging, but not as fast as the maximum rotational speed reported previously for isolated single $F_1$ complexes uncoupled from the membrane-embedded $F_o$ complex. Further application of ABEL trap measurements should help resolve the mechanistic causes of such fluctuating rates of subunit rotation.

# 1. INTRODUCTION

The synthesis of adenosine triphosphate (ATP) in most living cells is catalyzed by $F_oF_1$-ATP synthases. These enzymes are localized in the plasma membrane of bacteria, the thylakoid membranes in plant chloroplasts, and in the inner mitochondrial membrane of eukaryotic cells [1]. For $F_oF_1$-ATP synthase from *Escherichia coli* ($EF_oF_1$ in the following) and most other organisms, the driving force for ATP synthesis is the proton motive force (*pmf*), which is composed of both the electrochemical membrane potential ($\Delta\psi$) and the difference in proton concentrations across the plasma membrane ($\Delta$pH) [2, 3]. Maximum ATP synthesis rates at very high *pmf* were determined between 100 ATP·s$^{-1}$ for purified $EF_oF_1$ reconstituted in artificial liposomes at 23°C [4] and 270 ATP·s$^{-1}$ for membrane vesicles using the native lipid-protein composition at 37°C [5]. However, under realistic conditions with a smaller *pmf* and a physiological ATP/ADP ratio, ATP synthesis rates were significantly lower, i.e., in the range of 1 to 10 ATP·s$^{-1}$ measured in native membrane vesicles [6-8].

$F_oF_1$-ATP synthases can also catalyzes the reverse chemical reaction, the hydrolysis of ATP to ADP and phosphate. Catalytic turnover numbers for ATP hydrolysis at room temperature have been reported in the range of 60 ATP·s$^{-1}$ for reconstituted $EF_oF_1$ in a liposome [9], 145 ATP·s$^{-1}$ for $EF_oF_1$ reconstituted in lipid nanodiscs [10], or 300 ATP·s$^{-1}$ for $EF_oF_1$ solubilized in detergent micelles [11]. The slower rate for $EF_oF_1$ in sealed liposomes is largely due to inhibition by the *pmf* generated by ATP hydrolysis-driven proton pumping across the lipid bilayer. Accordingly, addition of protonophores or uncouplers to dissipate the counteracting *pmf* increased the maximum ATP hydrolysis rates of reconstituted $F_oF_1$-ATP synthases in liposomes by factors of two [12] to five [9].

To maintain the required physiological ATP/ADP ratio in living cells, catalysis of the ATP hydrolysis reaction by $F_oF_1$-ATP synthases is strictly controlled. Two types of independent inhibitory mechanisms have been identified. For the *E. coli* and other bacterial $F_oF_1$-ATP synthases, regulation of ATP hydrolysis is achieved either by a large conformational change of the C-terminal domain of the ε subunit (εCTD) in $F_1$ (εCTD inhibition) [13, 14], or by tight binding of ADP to one catalytic site of $F_1$ (ADP inhibition) [15]. For $EF_oF_1$ in native membrane vesicles, about 50% of the enzymes were estimated to be inactivated due to εCTD inhibition and another 30% were inactive due to ADP inhibition [16]. Truncation mutations of the εCTD resulted in threefold increased ATP hydrolysis rates of reconstituted $EF_oF_1$ [11]. The *pmf* might reduce the prevalence of the εCTD-inhibited or the ADP-inhibited states. For example, pre-energization of $EF_oF_1$-liposomes with *pmf* in the presence of ADP and phosphate resulted in a transient ninefold increase in the rate of ATP hydrolysis immediately following dissipation of pmf [9]. The role of *pmf* for activating and controlling $F_oF_1$-ATP synthases from various organisms is currently an important open question to understand the sophisticated regulatory mechanisms of the enzymes [17].

The fundamental "*rotary mechanism*" for sequential catalysis on the three catalytic sites in the $F_1$ part of $F_oF_1$-ATP synthases was proposed by P. Boyer (see review [18]). First unequivocal evidence for this mechanism was achieved by x-ray crystallography structures of $F_1$ parts reported from J. Walker's group [19], followed by sophisticated biochemical [20] and spectroscopic experiments [21]. However, video-microscopy provided direct observation of the unidirectionally rotating central γ-subunit in a single $F_1$ complex driven by ATP hydrolysis [22]. Functional surface attachment of the catalytic $F_1$ domain inaugurated single-molecule enzymology. Recording a light-scattering or a fluorescent marker on the rotating γ-subunit of $F_1$-ATPase revealed substrate-dependent kinetics of the catalytic rotational steps [23]. Kinetic heterogeneities were unraveled, either by comparing mean rates of individual $F_1$-ATPases (called static disorder [24]) or within the time trajectory of a single $F_1$-ATPase (called dynamic disorder [24]). Novel mechanistic insights include the substeps for substrate binding and product release at distinct rotary angles [25], back steps of the rotating γ-subunit [26], temperature dependence of rotation [27], and inhibition by small molecules including ADP [28] and by the εCTD [29]. Comparison of $F_1$ fragments from different species showed significant differences between the rotary substeps of the γ-subunit in bacterial and in mitochondrial $F_1$-ATPases [30-32]. To resolve an angular dependency for the distinct catalytic processes of ATP binding, hydrolysis and release of ADP and phosphate, very high spatial and temporal resolution is required. This is achieved by the combination of dark-field microscopy with high-speed cameras and using backscattered laser light from 40 to 80 nm gold particles at rotary subunits [25, 29]. Localization with nanometer precision is feasible at camera frame rates of 100 kHz.

Similarly, mechanistic insights were gained for the entire $F_oF_1$-ATP synthase. Using light-scattering or fluorescent markers on the ring of c-subunits enabled monitoring of subunit rotation in the $F_o$ part. $EF_oF_1$ in detergent or in lipid nanodiscs were bound to the surface *via* the static subunits of the $F_1$ part or *via* the rotating c-ring [10, 33]. ATP hydrolysis in $F_1$ powered the c-ring rotation in the absence of a *pmf*. However, gold nanoparticles are far too large to be used for monitoring the internal rotary movements of the γ- or ε-subunits in $F_oF_1$ [34]. Rotation markers must be as small as possible, i.e., single fluorophores. For example, ATP synthesis of single liposome-reconstituted *P. modestum* $F_oF_1$ was studied with a [$Na^+$] motive force [35]. Subunit rotation rates were comparable to catalytic rates from biochemical assays. However, individual rotary steps of the surface-attached $F_oF_1$ were not resolvable due to a low molecular brightness of the Cy3 fluorophore attached to the c-ring.

Especially for applying a *pmf*, confocal single-molecule Förster resonance energy transfer (smFRET) in solution was established to analyze the rotary steps of the $\gamma$-, $\varepsilon$- and *c*-subunits of $EF_oF_1$ in liposomes diffusing freely in solution [36]. One fluorophore was specifically bound to one of the rotating subunits and another fluorophore to a static subunit of $EF_oF_1$. Three rotary orientations of the $\gamma$- or $\varepsilon$-subunit corresponded to three distinct FRET efficiencies in inactive $EF_oF_1$. ATP synthesis was monitored in the presence of a high *pmf* [37]. Catalytically active enzymes provided stepwise FRET efficiency changes in smFRET time traces. Accordingly, rotation of the $\gamma$- and the $\varepsilon$-subunit [38] in the catalytic $F_1$ domain occurred in three steps and in opposite direction for ATP synthesis compared to ATP hydrolysis. Proton-driven *c*-ring rotation was a ten-step movement [39]. Moreover, catalytic turnover for each of the three binding sites was slightly different causing a "*kinetic limping*" [40].

ATP synthesis rates averaged from single $EF_oF_1$ were found in agreement with rates from biochemical assays of the same proteoliposomes. i.e., mean single-enzyme ATP synthesis rates were ~20 ATP·s$^{-1}$. Mean subunit rotation-based ATP hydrolysis rates were ~70 ATP·s$^{-1}$. However, smFRET experiments selected only active $EF_oF_1$ while biochemical assays comprised at least 80% inactive enzymes as well [16]. Why are the catalytic rates obtained from single active $EF_oF_1$ not five times faster than rates from biochemical assays? Using surface-immobilized $EF_oF_1$ in liposomes showed significantly slowed turnover in preliminary confocal smFRET imaging experiments [41]. Geometrical constraints by the liposome on a surface might strongly alter local concentrations of substrate and products for an $EF_oF_1$ diffusing towards the glass-liposome interface within the lipid bilayer.

In solution, mean observation times of 20 to 100 ms limited the detection of full rotations in single FRET-labeled $EF_oF_1$ to less than 4 rounds (3 ATP per round) of hydrolysis or synthesis. Time traces were too short to study slower conformational processes such as subunit rotation at lower ATP concentrations. Neither static nor dynamic disorder of individual $EF_oF_1$ turnover could be assessed. A solution to these limitations is the Anti-Brownian ELectrokinetic trap (ABEL trap) developed by A. E. Cohen and W. E. Moerner [42-46], . The confocal ABEL trap comprises a flat microfluidic chip, a fast laser focus pattern and a synchronized feedback system. This trap is designed to push a fluorescent nanoparticle to the center of the laser focus pattern in real time using matching electric field gradients. After a diffusing single object is trapped, it is hold in place without surface interference [47]. The ABEL trap allows to compare and optimize fluorophores for smFRET [48]. Here, observation times of single active $EF_oF_1$ in isotropic solution were prolonged by the ABEL trap. Using smFRET as the readout, up to 100 full subunit rotations in individual $EF_oF_1$ were recorded at high ATP concentrations. With a resolution of 1 ms per smFRET time bin, single $EF_oF_1$ turnover was analyzed for a range of ATP concentrations. We found that the fastest individual turnover of $EF_oF_1$ in a liposome could exceed 500 ATP·s$^{-1}$ in the absence of a *pmf*. Individual kinetic rates were broadly distributed and varied by factors > 10. Dynamic disorder was found within smFRET time traces. The buildup of a *pmf* limited ATP hydrolysis and contributed to the static disorder. We anticipate that smFRET in the ABEL trap may become a key approach for unravelling the single-molecule enzymology of $F_oF_1$-ATP synthases driven by a *pmf*.

## 2. EXPERIMENTAL METHODS

**Reconstituted FRET-labeled $F_o F_1$-ATP synthase**. For smFRET recordings in the ABEL trap we prepared a specifically labeled $EF_o F_1$ in two steps. First, we purified the $F_1$ part of $F_o F_1$-ATP synthase from *E. coli* carrying a cysteine mutation in the rotating ε-subunit at residue 56 (εH56C) [38]. Cy3B-maleimide was attached to εH56C with a labeling efficiency of 74% as the FRET donor. Secondly, we purified a different $EF_o F_1$ mutant with a cysteine introduced at the C-terminus of the non-rotating *a*-subunit (*a*CT) [49]. Alexa Fluor 647-maleimide was attached to *a*CT with a labeling efficiency of 32% as the FRET acceptor. This labeled $EF_o F_1$ was reconstituted into pre-formed liposomes using an estimated enzyme-to-liposome ratio of 1 to 4 [49]. Liposomes contained a 5% portion of negatively charged lipids. The $F_1$ part was removed quantitatively using buffer without $Mg^{2+}$ as done before [36]. Afterwards, the Cy3B-labeled $F_1$ was reattached to the $F_o$ part in liposomes. Excess, unbound Cy3B-labeled $F_1$ was removed by ultracentrifugation procedures [36]. Liposomes with one $EF_o F_1$ (proteoliposome in the following) had a mean diameter of about 120 nm, as published [38]. The FRET-labeled reconstituted $F_o F_1$-ATP synthase was fully functional with a mean ATP synthesis rate of 16 ±2 ATP·s⁻¹, i.e. in good agreement with similarly prepared FRET-labeled $F_o F_1$-ATP synthases [50]. A brief description of the purification procedures, labeling, reconstitution and catalytic activities of the FRET-labeled $F_o F_1$-ATP synthase is given in the Supporting Information online (Figures S1 to S6).

According to recent cryoEM structures of εCTD-inhibited *E. coli* $F_o F_1$-ATP synthase [51], the 120°-stepped rotation of the ε-subunit with respect to the static *a*-subunit will result in three distances between the Cα atoms of εH56C and *a*CT (Figure 1 A, B). Two short distances are 4.3 nm (εH56C position "$H_2$", Figure 1 C) or 4.5 nm (εH56C position "$H_1$", Figure 1 C), respectively, and a long distance is 7.5 nm ($\varepsilon$H56C position "L", Figure 1 C). In order to estimate FRET efficiencies corresponding to L, $H_1$ and $H_2$, we used a Förster radius $R_0$=7.2 nm for the FRET pair Cy3B and Alexa Fluor 647 and fluorescence quantum yields $\phi_{(Cy3B)}$=0.67 and $\phi_{(Alexa\ Fluor\ 647)}$=0.33 (data from the FPbase FRET calculator at https://www.fpbase.org/fret). We considered an additional mean linker length of 0.5 nm for the fluorophores. Accordingly, two significantly different FRET efficiencies ($E_{FRET}$) are expected to be distinguishable in time traces of single $EF_oF_1$ during ATP-driven $\varepsilon$-subunit rotation, i.e., $E_{FRET}$=0.35 for L (or "*medium FRET*") and $E_{FRET}$=0.90 for $H_1$ or $E_{FRET}$=0.92 for $H_2$, respectively. Both $H_1$ and $H_2$ are "*high FRET*" and are not distinguishable with our experimental setup (see also [47]). We simulated FRET traces of single $EF_oF_1$ in an ABEL trap and confirmed that $H_1$ and $H_2$ cannot be discriminated, both for short and for long dwell times (for instance see Figure 1 E, simulated with more distinct FRET efficiencies $E_{FRET}$=0.80 for $H_1$ and $E_{FRET}$=0.90 for $H_2$).

Reassembly of Cy3B-labeled $F_1$ to the Alexa Fluor 647-labeled $F_o$ domain in a liposome yielded a fully functional $F_oF_1$-ATP synthase. The reconstituted enzymes synthesized ATP from ADP and phosphate with a mean catalytic rate of 16 ATP·s$^{-1}$ (Supporting Information, Figure S6). $EF_oF_1$ activity was in good agreement with our previous results of rebinding $F_1$ onto $F_o$ domains in liposomes [38, 52].

**ABEL trap for smFRET**. The confocal ABEL trap microscope for smFRET measurements was built according to the published setup from the A. E. Cohen group [45]. The scheme of our setup is shown in Figure 2 A and an image is shown in the Supporting Information (Figure S7). Briefly, we used a 532 nm cw laser and two electro-optical beam deflectors (EOD$_{x, y}$) to move the laser focus in a predefined pattern within microseconds. The pattern comprised 32 focus positions (Figure 2

B) and covered a 2.34 x 2.34 $\mu m^2$ area in the focal plane (see color image Figure 2 C). This so-called 32-point knight tour pattern ensured a very homogeneous spatial and temporal illumination [46]. The beam waist of the laser focus was 0.6 $\mu m$, as measured from the intensity maximum to the $1/e^2$ value. The distance between adjacent focus positions was 0.47 $\mu m$, ensuring an overlap between subsequent knight tour positions. The combination of a 300 $\mu m$ pinhole (PH in Figure 2 A) in the detection pathway with the 60x oil immersion objective with numerical aperture 1.42 allowed to detect all emitted photons from any position of the laser-excited area.

The completed knight tour was repeated at 5 or 7 kHz rate and was controlled by a field-programmable gate array (FPGA) card. The published FPGA-based ABEL trap software from the A. E. Cohen group [45] was slightly modified for utilizing up to four detectors in our setup [53]. Here we used the combined photon counts from two single photon counting avalanche photodiodes (APD), i.e., from FRET donor and FRET acceptor channel. The FPGA software directed the laser focus stepwise and allocated the recorded photons to each focus position. Once a fluorescent $EF_oF_1$ in a liposome entered the trapping area and started to emit photons, the software estimated the probable position of this proteoliposome. Next the software predicted the direction and speed of diffusion based on a previous position estimation, and generated feed-back voltages to four Pt-electrodes. Symmetric electric potentials with opposite signs were applied to each pair of electrodes for x- or y-direction (Figure 2 D). Electrodes were inserted into the sample chamber through holes in the PDMS block (Figure 2 E). Electrodes were in contact with the buffer solution so that electrophoretic and electroosmotic forces could push the negatively charged proteoliposome to the center of the focus pattern. An electric field strength of ~1.2 mV / $\mu m$ had been calculated for this sample chamber design [54]. It was linear across the trap region and up to the ends of the flat channel (Figure 2 D). All photons were recorded in parallel by a time-correlated single photon counting card (TCSPC) and were used for the smFRET analysis. Further details of the confocal ABEL trap setup are given in the Supporting Information.

**Microfluidic PDMS sample chamber**. To confine proteoliposome diffusion in two dimensions (x and y), a microfluidic ABEL trap chip was made from structured PDMS bonded to cover glass after plasma etching (Figure 2 E). Preparation details for PDMS-on-glass ABEL trap chips are given in the Supporting Information (Figure S8). The height in the trap region was 1 μm, limiting the smFRET detection volume in the ABEL trap region of the chip to ~9 fL. The total volume of the chip including the 80 μm deep channels for the electrodes was ~10 μl. The design of the PDMS chip was adopted from previous publications [44, 54]. To confirm the size of the detection volume, diffusion times of a diluted rhodamine 6 G solution (R6G) in $H_2O$ were measured by fluorescence correlation spectroscopy (FCS). With a fixed laser focus position, the diffusion time of R6G was $\tau_D$=0.35 ms. Applying the 32-point laser pattern in the absence of feed-back to the electrodes increased the mean diffusion time 5-fold to $\tau_D$~1.7 ms (Supporting Information, Figure S9). This is in good agreement with the corresponding increase of the laser-excited area (Figure 2 B). Manual z-drift correction of the laser focus was achieved by lowering the microscope objective. Refocusing was necessary because of a continuing z-drift in the range of 1 μm in 5 to 10 minutes (see also [56]).

**Analysis of smFRET traces**. Proteoliposomes were diluted to a concentration of 20 to 50 pM labeled $EF_oF_1$ and filled into the freshly prepared ABEL trap sample chamber. FRET donor photons were detected in the spectral range between 544 nm and 620 nm and FRET acceptor photons for wavelengths $\lambda > 647$ nm. Using time-correlated single photon counting (TCSPC) electronics with a time resolution of 164 ps, time traces were recorded for 25 min by multiplexing the APD signals for TCSPC in parallel to the FPGA input for ABEL trapping. The software "Burst Analyzer" [57] was used to visualize the fluorescence intensity time traces of FRET donor Cy3B and FRET acceptor Alexa Fluor 647 as well as the sum intensity of both detection channels. When a proteoliposome entered the trapping area, immediate trapping was characterized by a stepwise rise of the combined fluorescence count rate (40 to 50 counts per ms above background) within a single time bin of 1 ms. Photon bursts with lower or with higher total count rates were omitted from subsequent smFRET analyses. Loss from the trap or FRET donor photobleaching was accompanied by a stepwise drop of the total count rate to the background level. Photon bursts were manually marked using stepwise rise and drop of the fluorescence count rate as the criterion. Background had to be subtracted for each photon burst individually because the background decreased on both channels in a time-dependent manner. Decreasing background was caused by bleaching of impurities from within the PDMS block and from the cover glass, for example, due to adsorbed fluorophores or adsorbed proteoliposomes within the trapping area. For each photon burst, the start time within the recording, its duration, and the number of FRET state fluctuations were noted. A full 360° rotation of the ε-subunit with respect to the $a$-subunit of $F_oF_1$-ATP synthase comprised one "medium FRET" state (L) plus one "high FRET" state ($H_1$ or $H_2$), i.e., each pair of FRET states corresponded to the supposed hydrolysis of 3 ATP molecules by the enzyme. We simulated smFRET traces using Monte Carlo simulations [39]. Applying similar dimensions and photophysical FRET parameters in the simulations revealed the accuracy of the quick manual assignment approach for different turnover values.

**Photophysics of FRET donor and FRET acceptor fluorophores bound to $F_oF_1$-ATP synthase**. To choose the optimal FRET donor for 532 nm excitation, we compared the photophysical behavior of our previously used FRET donor tetramethylrhodamine (TMR) [36-38] with Cy3B or Atto R6G bound to $EF_oF_1$. Reconstituted TMR-labeled $EF_oF_1$ exhibited a mean brightness of 14 counts per ms with a full width at half maximum (FWHM) of the brightness distribution of 4 counts per ms (Supporting Information, Figure S11). The trapping times of the proteoliposomes varied. The distribution of photon burst durations had a maximum around 200 ms, with a few bursts lasting up to 2 seconds. Fitting the distribution with an exponential decay function yielded an average duration of 350 ms. The reconstituted Cy3B-labeled $F_oF_1$-ATP synthase exhibited a much higher mean brightness of 34 counts per ms, with a FWHM of the brightness distribution of 5 counts per ms (Supporting Information, Figure S11). The distribution of photon burst durations was different, with more prolonged trapping times and a larger average duration of 750 ms. Background was comparable to the measurements of the TMR-labeled enzyme, i.e., decreased from 15 counts per ms at the beginning to 10 counts per ms within 5 min. Alternatively we examined AttoR6G as a FRET donor. AttoR6G-labeled reconstituted $EF_oF_1$ exhibited a mean brightness of 41 counts per ms, i.e., higher than Cy3B or TMR. The intensity distribution was asymmetrically broadened with a FWHM of 8 counts per ms (Supporting Information, Figure S11). The duration distribution was comparable to TMR-labeled $F_oF_1$-ATP synthases, with few bursts lasting up to 2.5 seconds, yielding an average duration of 370 ms for trapped proteoliposomes.

Therefore, we chose Cy3B on the ε-subunit of the $F_1$ domain as the FRET donor because of its relatively high single-molecule brightness, the narrow and symmetric brightness distribution confirming a lack of protein-induced fluorescence changes [58] and the significantly prolonged residence times of Cy3B-labeled $EF_oF_1$ in the ABEL trap. Selecting Alexa Fluor 647 as the FRET acceptor provided a large Förster radius, $R_0$, and allowed to test the use of a triplet quencher in order to increase the brightness or to diminish blinking of both fluorophores [59].

## 3. RESULTS AND DISCUSSION

ATP-driven subunit rotation within individual *E. coli* $F_oF_1$-ATP synthases can be detected by single-molecule FRET in solution [36] (Figure 1). Therefore we labeled one genetically introduced cysteine in the rotary ε-subunit of the $F_1$ domain with Cy3B and another in the static a-subunit of the $F_o$ domain with Alexa Fluor 647. According to recent structures [51] of inhibited $EF_oF_1$, we expected one large distance between the two markers (i.e., FRET state L for one rotary position of ε) and two indistinguishable short distances (FRET states $H_1$ and $H_2$). Sequential changes of FRET states →L→$H_1$→$H_2$→L→ were anticipated for ε-subunit rotation during ATP hydrolysis [38]. For smFRET recordings we utilized a confocal microscope setup called an Anti-Brownian ELectrokinetic trap [42] (ABEL trap, Figure 1 D, see Experimental Methods above). In general, the ABEL trap localized the spatial position of the FRET-labeled enzyme in solution and pushed it back in real time to the center of the μm-sized trapping region (Figure 2). Extended smFRET observation times of single $EF_oF_1$ in the membrane of ~120 nm-sized liposomes allowed to vary ATP concentrations from 1 mM down to 5 μM. Besides analyzing fluctuating and stochastic turnover, we aimed to confirm whether the buildup of a proton motive force (*pmf*) limits ATP hydrolysis rates differently for distinct individual $EF_oF_1$-liposomes.

**Two distinct FRET states in the presence of AMPPNP**. At first we ascertained the distinct FRET states of the rotary ε-subunit during catalysis. AMPPNP is a non-hydrolyzable derivative of ATP that binds with similar affinity to $EF_oF_1$ but stalls functional subunit rotation. Addition of AMPPNP results in a rotor orientation in one of three angular positions with respect to the *a*-subunit of the peripheral stalk. For single-molecule measurements in the ABEL trap, the proteoliposome suspension was diluted to an enzyme concentration of approximately 20 pM in buffer (10 mM tricine, 10 mM succinate, 0.3 mM KCl, pH 8.0) containing 1.25 mM $Mg^{2+}$ and 1 mM AMPPNP.

Figure 3 shows example measurements with individual trapped $F_oF_1$-liposomes. The actual mean background photon counts on both detector channels had been subtracted and FRET was calculated using the proximity factor $P=I_{acceptor}/(I_{donor} + I_{acceptor})$ instead of a fully corrected FRET efficiency[60]. Most photon bursts (i.e., 84% of 1593) were not exhibiting a FRET acceptor signal and were called "donor only"-labeled enzymes (see photon bursts with numbers I, V, VI, VII in Figure 3A). Due to spectral crosstalk of Cy3B emission into the FRET acceptor channel, the mean apparent proximity factors of the "donor only" photon bursts were $P=0.15\pm0.05$. However, enzymes with distinct P values of 0.4 (medium FRET, photon burst II) or $P>0.8$ (high FRET, photon bursts III, IV) were also found. In photon burst IV, the FRET acceptor Alexa Fluor 647 (red trace) photobleached after 90 ms but the proteoliposome remained trapped due to the FRET donor signal (green trace). Figure 3 B shows another enzyme with mean $P\sim0.4$, held in solution by the ABELtrap for 4.3 seconds. Figure 3 C shows an enzyme with mean $P\sim0.9$ held for 1.4 seconds, with FRET acceptor photobleaching near the end of this photon burst.

The histogram of P values from all FRET-labeled $EF_oF_1$ in Figure 3 D indicated two subpopulations with comparable occurrence. The two FRET states were well separated. The $P\sim0.4$ state was associated with higher total photon counts for the sum of FRET donor and acceptor signals ($\sim37$ counts per ms in Figure 3 E) in comparison to the $P\sim0.85$ state with $\sim20$ counts per ms. The "donor only" state with apparent $P\sim0.15$ exhibited about 45 counts per ms as added intensities from both detection channels (Supporting Information, Figure S12). Summed photon signals were depicted as black traces in Figures 3 A-C. The FRET state duration depended on the proximity factor, i.e., FRET states with medium P values lasted longer (Figure 3 F) and with similar lengths (up to 5 seconds) as "donor only" photon bursts (Supporting Information, Figure S12). FRET states with high P values were found lasting up to 2 seconds. FRET acceptor photobleaching was observed for both medium and high P. The difference in FRET state duration could be caused by faster Alexa Fluor 647 photobleaching in the high FRET state or by predominant loss of the high FRET state from the trap due to lower total photon count rates and, accordingly, an unfavorable signal-to-background ratio.

To enhance the photostability of the cyanine dye Alexa Fluor 647, we tested an addition of the antioxidant 6-hydroxy-2,5,7,8-tetramethylchroman-2-carboxylic acid (*trolox*, [61]) to the buffer in the presence of 1 mM AMPPNP. In the presence of 2 mM *trolox*, AMPPNP-stalled $EF_oF_1$ showed the same two proximity factor states P~0.4 and P~0.85. About 80% of $EF_oF_1$ were found as "donor only"-marked (Supporting Information, Figure S12). The apparent proximity factor of the "donor only" photon bursts was P=0.15±0.05 (Supporting Information, Figure S12).

Slightly different brightnesses of individual FRET donor Cy3B were recognized in the presence of *trolox*. Cy3B-labeled $F_oF_1$ ATP synthases emitted between 35 and 45 counts per ms, resulting in a broadened distribution of summed-intensity-*versus*-P values (Supporting Information, Figure S12). However, the apparent proximity factor remained the same, i.e., FRET donor brightness did not affect the P values. Importantly, the distribution of FRET state durations did not display a significant prolongation but was comparable to that for photon bursts in the absence of the antioxidant *trolox*. Adding 2 mM *trolox* yielded no obvious photophysical improvement for the smFRET recording in the ABEL trap.

**Two fluctuating FRET states in the presence of 1 mM ATP**. To monitor catalytic turnover of $EF_oF_1$ at saturating substrate concentration, 1 mM ATP and 1.25 mM $Mg^{2+}$ were added to the proteoliposome suspension. Although most $EF_oF_1$ showed a constant FRET efficiency with proximity factors of either P~0.4 or P~0.85, we also observed many photon bursts with fast oscillating FRET levels (Figure 4). P fluctuations were caused by anticorrelated changes of FRET donor and acceptor intensities. As these fluctuations were not observed in the absence of ATP, we attributed these P fluctuations to ε-subunit rotation driven by ATP hydrolysis. P fluctuations were analyzed as pairs of one medium FRET level (P<0.5) plus one high FRET level (P>0.8) in the time trace. Each FRET pair was interpreted to represent a full 360° rotation of the ε-subunit because the two different high FRET level $H_1$ and $H_2$ were expected to be indistinguishable. Accordingly, the duration of one pair of FRET levels was correlated with three hydrolyzed ATP molecules in that time span.

Manual assignment of beginning and end of FRET fluctuations within a photon burst and counting the number of FRET pairs within the photon burst allowed to assign a mean catalytic rate, or ATP hydrolysis rate, respectively, for each individual FRET-labeled $F_oF_1$-ATP synthase hold by the ABEL trap. For example, the enzyme shown in Figure 4 A exhibited 74 FRET fluctuations (or pairs of FRET level) within a period of 1118 ms before the FRET acceptor Alexa Fluor 647 photobleached. Afterwards the enzyme remained as a "donor only"-labeled $EF_oF_1$ trapped for another 327 ms. The number of full rotations per period corresponded to an average of 15.1 ms per 3 hydrolyzed ATP molecules and a mean ATP hydrolysis rate of 199 ATP·$s^{-1}$. This $EF_oF_1$ was recorded 6 min after starting the ABEL trap measurement.

Figure 4 B depicts signals from another $EF_oF_1$ that were recorded 11.4 min after starting the ABEL trap measurement. Here, the FRET fluctuations occurred significantly slower and with a clear stepwise change between the medium and high FRET levels. Within 1250 ms duration, 17 FRET level pairs were assigned, corresponding to 73.5 ms per full rotation and a mean ATP hydrolysis rate of 41 ATP·$s^{-1}$. After FRET acceptor photobleaching the "donor only"-labeled $EF_oF_1$ remained trapped for another 386 ms.

The rotational speed varied from enzyme to enzyme. However, some trapped $EF_oF_1$ also exhibited apparent changes of the catalytic rate during the period of the FRET fluctuations. In Figure 4 C the mean catalytic rate was ~60 ATP·s$^{-1}$. The first FRET fluctuation period for this $EF_oF_1$ was slower with a mean of 36 ATP·s$^{-1}$ during 1264 ms and was followed by a faster period with 108 ATP·s$^{-1}$ for 417 ms. This $EF_oF_1$ was recorded 9 min after starting the ABEL trap measurement. A second example of dynamic changes of the catalytic rate of $EF_oF_1$ is shown in Figure 4 D. In the first part, i.e., for 342 ms, the rate was very fast with 347 ATP·s$^{-1}$, then slowed to 64 ATP·s$^{-1}$ for 235 ms, and accelerated in the second half of the photon burst to 190 ATP·s$^{-1}$ before the FRET acceptor photobleached. This $EF_oF_1$ was recorded 5.3 min after starting the ABEL trap measurement. In addition, $EF_oF_1$ with much faster FRET fluctuations were observed. In Figure 4 E a mean ATP hydrolysis rate 356 ATP·s$^{-1}$ was calculated from 118 rotations in 994 ms. This $EF_oF_1$ was recorded 4 min after starting the ABEL trap measurement.

**Distribution of catalytic rates from individual enzymes in the presence of 1 mM ATP**. The wide range of ATP-driven rotation rates in single $F_oF_1$-ATP synthases was analyzed with respect to the start of the smFRET recording with ABEL trapping. Recording started about 2 to 3 minutes after addition of ATP to the proteoliposome suspension. The delay was caused by filling of the ABEL trap chamber, inserting the electrodes and aligning the trapping region on the microscope stage. A series of five independent measurements were combined in Figure 5 A, with each recording spanning the time course of 25 min. No obvious trends in the distribution of catalytic rates were found. From start to end, the individual mean ATP hydrolysis rates were scattered between 20 ATP·s$^{-1}$ to 400 ATP·s$^{-1}$ in the presence of 1 mM ATP (Figure 5 A, left diagram). The rate distribution had a broad maximum in the range of 100 to 200 ATP·s$^{-1}$ (Figure 5 A, middle diagram). No correlation could be found between the number of rotations per photon burst and the associated ATP hydrolysis rate (Figure 5 A, right diagram), excluding a bias from $EF_oF_1$ with a longer residence times in the ABEL trap and, therefore, exhibiting more rotations. The ATP hydrolysis rate distribution was essentially determined by photon bursts with less than 20 rotations. Only 2 from a total number of 289 single $F_oF_1$-ATP synthases were recorded exhibiting more than 100 full rotations. Different rotational speed or different catalytic rates from enzyme to enzyme is called *static disorder* in single-molecule enzymology, whereas the variation of catalytic speed including a stop-and-go behavior of a single enzyme is called *dynamic disorder* [62].

**ATP-concentration dependent catalytic rates from 5 μM to 1 mM ATP.** We repeated the measurements of individual ATP hydrolysis rates using lower ATP concentrations, i.e., in the presence of either 100 μM ATP, 40 μM ATP, 20 μM ATP or 5 μM ATP. Examples of photon bursts at all ATP concentrations are given in the Supporting Information (Figures S13 and S14). Fast and slow catalytic turnovers were observed at each ATP concentration, as calculated from the pairs of medium and high FRET levels. The transition between the P~0.4 and P~0.85 levels occurred fast in the range of one to a few ms. However, at 5 μM ATP these transitions became longer in some photon bursts. P transitions occurred either in a continuous, "ramp-like" behavior, or stepwise with the high FRET level split into two distinct FRET levels at P>0.85 and at 0.6<P<0.8 (Supporting Information, Figure S14).

The single-enzyme rate distributions are summarized in Figure 5. For each ATP concentration the catalytic rates varied substantially by a factor of 10 at 100 μM ATP (36 to 360 ATP·s$^{-1}$ for 175 enzymes), 13 at 40 μm ATP (27 to 346 ATP·s$^{-1}$ for 184 enzymes), 16 at 20 μM ATP (15 to 238 ATP·s$^{-1}$ for 112 enzymes), and a factor of 8 at 5 μM ATP (14 to 106 ATP·s$^{-1}$ for 89 enzymes). No time-dependent decrease of catalytic rates was observed even for the low ATP concentrations, i.e., the initial substrate concentrations were not altered significantly by the enzymatic turnover within the 25 min recording time. However, at 20 μM and at 5 μM ATP, the average ATP hydrolysis rates of individual $F_oF_1$-ATP synthases were smaller than those at the higher ATP concentrations, as expected from basic enzyme kinetics and biochemical assays.

Consistent with the ATP-dependent decrease of mean catalytic turnover, the maximum number of rotations per FRET fluctuation period decreased with lower substrate concentration from about 90 rotations at 100 μM ATP to around 15 at 5 μM ATP (right diagrams in Figure 5). The distributions of FRET fluctuation durations of single $EF_oF_1$ were similar for all ATP concentrations with upper limits between 1 to 1.5 seconds (Supporting Information, Figure S15).

**Dissipating the *pmf* with ionophores.** A possible cause for the broad distributions of catalytic rates could be a variable *pmf* for each proteoliposome. Due to the stochastic ATP-driven proton pumping by $F_oF_1$-ATP synthase, the actual ΔpH and/or membrane potential might counteract ATP hydrolysis differently in individual proteoliposomes. The liposomes have slightly different sizes. Therefore, the addition of ionophores and uncouplers was investigated (Figure 6). At first, we measured the effect of the protonophore carbonyl cyanide m-chlorophenyl hydrazone (CCCP, 50 μM) in the presence of 100 μM ATP. FRET-labeled $EF_oF_1$ were observed with fast transitions between the medium and high FRET levels. The $EF_oF_1$ in Figure 6 A hydrolyzed ATP with a rate of 418 ATP·s$^{-1}$. The ATP hydrolysis rates for different individual enzymes varied between 112 to 537 ATP·s$^{-1}$ (factor of 5, Figure 6 B). The average ATP hydrolysis rate from all enzymes was 354 ATP·s$^{-1}$, i.e., the mean turnover was 2.5-fold faster than for 100 μM ATP (144 ATP·s$^{-1}$, Figure 5 B) without the protonophore CCCP. Apparently, a slightly slower mean could be estimated towards the end of the 25 min recording time.

Subsequently we compared the dissipation of a *pmf* by the combination of valinomycin acting as a $K^+$ carrier across the lipid bilayer and nigericin acting as an antiporter for $H^+$ and $K^+$. Valinomycin and nigericin were added at 1 µM concentration each in the presence of 100 µM ATP. Fast transitions between the medium and high FRET level were observed in FRET-labeled $EF_oF_1$. The enzyme shown in Figure 6 C exhibited 117 full rotations in 1027 ms, i.e., a catalytic rate of 342 $ATP \cdot s^{-1}$.

Individual ATP hydrolysis rates varied between 111 to 496 $ATP \cdot s^{-1}$ by factor 5 (Figure 6 D). The average ATP hydrolysis rate from all enzymes was 295 $ATP \cdot s^{-1}$, i.e., the mean turnover was twofold faster than in the presence of 100 µM ATP only. A twofold increase of average ATP hydrolysis rates in biochemical assays in the presence of valinomycin and nigericin had been reported recently using $EF_oF_1$ reconstituted in liposomes [12]. However, a concentration of 300 µM $K^+$ in the ABEL trap buffer here might be insufficient to dissipate a *pmf* quickly by the combination of $K^+$ carrier valinomycin and $H^+/K^+$ antiporter nigericin.

The distributions of all individual $F_oF_1$-ATP synthase turnover numbers are combined in the box-and-whisker plots in Figure 6 E. As defined, the lower limit of each box was constructed from the 25% threshold (1st quartile) of the ATP hydrolysis data and the upper limit from the 75% threshold (3rd quartile). The median of all the rate distributions, i.e., the horizontal line within the interquartile range, was almost identical to the mean value of each set (dot in the box). The data sets recorded in the presence of 100 µM ATP showed that the interquartile ranges of turnover numbers for the measurements without (green box) and with uncouplers CCCP (dark yellow box) or valinomycin plus nigericin (yellow box) did not overlap. Accordingly, single-enzyme catalytic rates were significantly higher in the presence of uncouplers. These distinct distributions confirmed that the buildup of a ΔpH and an electrochemical membrane potential by ATP-driven proton pumping into the proteoliposomes reduced ATP turnover of $EF_oF_1$. However, the ranges defined by the whiskers (i.e., lines extending the box) of the distributions in the presence of uncouplers still overlap with the interquartile range for 100 µM ATP (green box) revealing that dissipating the *pmf* by uncouplers was not sufficient to minimize the wide variability of the rate distributions. Outliers (see single data points below and above the whiskers in Figure 6 E) were identified in each data set of ATP hydrolysis rates. Only a few $EF_oF_1$ exhibited very fast ATP turnover >300 ATP·s⁻¹ at 1 mM, 100 µM, or 40 µM ATP.

For the lower ATP concentrations 20 µM (cyan box) and 5 µM (purple box) their interquartile ranges did not overlap with the other boxes (red, green, blue), indicating that catalytic turnover on average was significantly slower. We tested the possibility to obtain kinetic parameters from the median values of the ATP-dependent distributions according to the Michaelis-Menten theory. Using Hanes-Woolf linearization [63] in Figure 6 F, the maximum ATP hydrolysis rate $V_{(max)}$ = 169 ATP·s⁻¹ was calculated from the slope of the fit (red line). The Michaelis-Menten constant $K_m$ = 17 µM was calculated from the intercept and from $V_{(max)}$. Alternatively, using Lineweaver-Burk linearization [64], values were calculated for $V_{(max)}$ = 147 ATP·s⁻¹ and $K_m$ = 12 µM (Supporting Information, Figure S16). These $V_{(max)}$ values of reconstituted $EF_oF_1$ were smaller than $V_{(max)}$ values of 651 or 516 ATP·s⁻¹ reported for single $EF_oF_1$ solubilized in detergent, attached to a glass surface and measured at room temperature using dark-field microscopy [65]. The smFRET-based $V_{(max)}$ and $K_m$ were in agreement with biochemical assays of *E. coli* $F_oF_1$-ATP synthase using native membrane vesicles ($V_{(max)}$~200 ATP·s⁻¹ [5], $V_{(max)}$ =336 ATP·s⁻¹ at 30°C [66]), purified $EF_oF_1$ in detergent ($V_{(max)}$ =285 ATP·s⁻¹ and $K_m$ =78 µM at 24°C [11], $V_{(max)}$ =217 ATP·s⁻¹ and $K_m$ =140 µM at 30°C [67], $V_{(max)}$ = 700 ATP·s⁻¹ with 10 mM ATP and 5 mM $Mg^{2+}$ at 37°C [68]), single $EF_oF_1$ reconstituted into lipid nanodiscs ($V_{(max)}$ = 140 ATP·s⁻¹ at 25°C [10]) or purified $EF_oF_1$ reconstituted into liposomes (v = 65 ATP·s$^{-1}$ at 100 μM ATP, increasing to v = 210 ATP·s$^{-1}$ in the presence of uncouplers valinomycin and nigericin [9, 12]).

$V_{(max)}$ obtained from Figure 6 F was about twice as fast as the mean turnover of ~70 ATP·s$^{-1}$ at 1 mM ATP reported previously in single-molecule FRET experiments with freely diffusing reconstituted $EF_oF_1$ in solution [38, 50, 69]. However, in those previous smFRET experiments, ATP turnover was determined differently. At first, dwell times of individual FRET levels were assigned, with a minimal duration of 2 to 5 ms for each dwell. The arbitrary pathway of freely diffusing $EF_oF_1$ through the confocal volume resulted in strong total fluorescence intensity fluctuations, i.e., strong fluctuations of the FRET efficiency for one assigned FRET level occurred as well. A bias to skip short intermediary dwells and to preferentially assign longer dwell times could not be excluded. Short observation times of freely diffusing proteoliposomes resulted in a few turnovers per enzyme only. Therefore all dwell times from different active $EF_oF_1$ had to be combined before fitting the dwell time histogram. For the actual ABEL trap data analysis, the turnover of each enzyme could be calculated separately based on many full rotations before sorting the individual rates into the ATP turnover histograms.

Alternatively, the faster ATP turnover of single $EF_oF_1$ in the ABEL trap could be caused by small local temperature rises due to the fast-switching feedback voltages. Voltages switched in less than 10 μs. Fluctuating electrokinetic forces might result in an additional active transport contribution for both substrate and products. For diffusion-limited kinetics, this could cause a higher ATP turnover of $EF_oF_1$ in the ABEL trap.

## 4. CONCLUSIONS

We recorded ATP hydrolytic turnover of individual *E. coli* $F_oF_1$-ATP synthase reconstituted with one enzyme per liposome. ATP concentrations varied from 5 µM to 1 mM ATP. We analyzed catalysis by confocal smFRET in solution and employed an ABEL trap to prolong single-enzyme observation times up to several seconds. With FRET donor Cy3B attached specifically to the rotating ε-subunit and FRET acceptor Alexa Fluor 647 attached to the static a-subunit., both fluorophores did not exhibit varying brightness after covalent attachment to the protein. A full rotation of the ε-subunit in $EF_oF_1$ correlated with three hydrolyzed ATP molecules. Using the proximity factor P as the measure of FRET efficiency, we exploited very distinct P values for the rotary positions of the ε-subunit. Full rotations were counted as pairs of minimum-maximum P values. Thus, time-consuming and error-prone efforts were avoided in assigning transition points precisely and in calculating dwell times for each individual FRET level. Held by the ABEL trap, most FRET-labeled $EF_oF_1$ exhibited less than 20 full rotations. Our straightforward counting approach followed previous subunit rotation measurements in single immobilized, highly-active $EF_1$ complexes [29].

**The maximum speed of ε-subunit rotation in reconstituted $F_oF_1$-ATP synthase and causes for static and dynamic disorder**. The maximum catalytic rates estimated for surface-attached single $F_1$ complexes (1350 ATP·s$^{-1}$ [70] or 2160 ATP·s$^{-1}$ [29] at room temperature) were not reached with reconstituted *E. coli* $F_oF_1$-ATP synthases in the ABEL trap. Undisturbed from any surface-related interference, the maximum ATP hydrolysis rate of a single $EF_oF_1$ in the presence of 1 mM ATP was 400 ATP·s$^{-1}$. However, under the same substrate conditions $EF_oF_1$ were recorded exhibiting a much slower turnover of 20 ATP·s$^{-1}$. The wide variation of individual catalytic rates continued during the recording time and the heterogeneities occurred for all ATP concentrations tested. This heterogeneity of catalytic activity from one enzyme to the next had been called *static disorder* [62], observable only by single-molecule enzymology. To elucidate the causes of catalytic heterogeneities between individual $EF_oF_1$ and within a turnover time trace (*dynamic disorder*), we reconsider the spatial and the temporal uniformness of the local environment of the $F_oF_1$-ATP synthase.

First, the reconstitution approach with re-attaching the FRET donor Cy3B-labeled $F_1$ domain back onto the FRET acceptor Alexa Fluor 647-labeled $F_o$ domain in liposomes allowed to circumvent analyzing any $F_oF_1$-ATP synthase with a 'wrong' outside-in orientation, because the $F_1$ domain can only bind to the correctly inserted $F_o$ domain. Because we analyzed ATP-driven FRET fluctuations only, all inactive enzymes were excluded, and their fraction did not influence the kinetic results.

Second, holding single proteoliposomes in solution by the ABEL trap allowed to record smFRET time traces of $EF_oF_1$ during ATP hydrolysis in an isotropic environment and for up to two seconds. The diameter of a proteoliposome in the range of ~120 nm was small with respect to the 1.1 μm height of the trapping region of the PDMS-on-glass sample chamber. Therefore, diffusion of the substrate ATP to the enzyme and diffusion of the products ADP and phosphate from the enzyme were not spatially restricted in the trap. During each 25 min recording time, a systematic change or an average slowdown of the turnover was not found. ATP consumption by $EF_oF_1$ diluted to less than 50 pM was negligible, so that the temporal uniformness of substrate and products for the catalytic turnover was ensured.

Third, one cause of *static disorder* for reconstituted $F_oF_1$-ATP synthase is the buildup of a *pmf* counteracting the ATP hydrolysis reaction. Adding protonophores or uncouplers revealed $EF_oF_1$ with a faster maximum ATP hydrolysis rate up to 537 ATP·s$^{-1}$, supporting that a *pmf* is limiting the catalytic rate. Adding saturating uncoupler concentrations did not result in more narrow turnover distributions, so that future smFRET rotation experiments might require EF$_o$F$_1$ reconstituted in a lipid nanodisc in order to prevent any residual *pmf* as the cause of *static disorder*.

Finally, the regulatory mechanism for ATP hydrolysis in EF$_o$F$_1$-ATP synthase, i.e., conformational changes of the εCTD to mechanically inhibit the motion of the γ, ε and *c*-ring rotor, might contribute to the *static disorder* as well as the *dynamic disorder*. Beyond the extended εCTD conformation [71] ("up") to block rotary catalysis and the compact active "down" conformation [72], intermediate and different extension states of the εCTD have been reported for the *E. coli* enzyme [49, 73], and the transition times between them are unknown. Depending on the time scales of conformational changes between the different ε-subunit states, slow changes could be related to the *static disorder* observed for different enzymes, and fast changes could explain the *dynamic disorder* within the time trace of individual catalytically active EF$_o$F$_1$. To unravel the role of εCTD conformations for the catalytic heterogeneities of EF$_o$F$_1$ in the future, we will analyze catalysis of εCTD truncation mutants by smFRET in the ABEL trap, for example the ε91-stop mutant with a complete removal of the εCTD [11] and the εΔ5 mutant with the last 5 amino acids removed [7]. Further, direct FRET monitoring of the εCTD conformation in single EF$_o$F$_1$ (as initiated in [49, 73]) could be applied in the ABEL trap to obtain the time constants of εCTD movements.

**Towards precise smFRET-based conformational analysis of EF$_o$F$_1$ in the ABEL trap.**
Accurate distance determination between two marker positions in EF$_o$F$_1$ requires corrections of FRET donor and acceptor intensities. Corrections have to compensate for the spectral transmission of the microscope objective, spectral properties of the optical filters, spectral sensitivity of the APD detectors, and fluorescence quantum yields of the fluorophores [74]. Photon counts of individual FRET-labeled $EF_oF_1$ are maximized by the ABEL trap [47]. High photon counts will allow to precisely identify distinct conformations of active $EF_oF_1$ during turnover. Thus smFRET-based $EF_oF_1$ conformations (i.e., active and with rotary substeps, ε-inhibited or ADP-inhibited, during ATP hydrolysis or synthesis, with small inhibitors bound, lipid-dependent, and *pmf*-dependent) can be compared with new structures and distributions of $EF_oF_1$ conformations based on cryoEM measurements [51, 75].

For example, most recent cryoEM structures provided evidence for an additional relative movement of the ε-subunit [76]. An $EF_oF_1$ subpopulation exhibited a rotary ε-subunit shift of about 60°, but the εCTD was found in a "down" conformation that allows activity. The shifted ε-subunit was interpreted as a conformation related to the ADP-inhibited state. As very rare events, we have found FRET-labeled $EF_oF_1$ with intermediate proximity factors P~0.6 that were long-lasting and almost static, which might represent these postulated novel ADP-inhibited conformations. Next these smFRET data have to be corrected and transformed to yield distances and rotary conformations. Unraveling these regulatory conformational dynamics will be crucial to fully understanding how this membrane enzyme works in both catalytic directions, i.e., synthesis and hydrolysis of ATP. This information should be accessible by single-molecule spectroscopy, especially by quantitative smFRET in isotropic solution to avoid surface-related artefacts, and the ABEL trap will become an essential tool to record long time trajectories of active $F_oF_1$-ATP synthase at work.

**Figure legends**

**Figure 1. A-C,** structure of *E. coli* $F_oF_1$-ATP synthase with cysteine mutations for labeling with FRET donor Cy3B on the rotating ε-subunit (green spheres) and FRET acceptor Alexa Fluor 647 on the static C-terminus of the *a*-subunit (red sphere). **A**, side view with highlighted rotor subunits ε (green), γ (cyan) and $c_{10}$-ring (blue) and all static subunits in grey, based on PDB structure 6OQV for the L orientation of ε and γ [51]. **B**, view from the bottom. **C**, assignment of FRET states L, $H_1$ (PDB structure 6OQW) and $H_2$ (PDB structure 6OQT). During ATP hydrolysis, ε-subunit rotation in counter-clockwise direction will cause a FRET state sequence →L→$H_1$→$H_2$→L→ [51]. **D**, a single active $EF_oF_1$ in a liposome is hold in place in solution by electrophoretic and electroosmotic forces in the Anti-Brownian ELectrokinetic (ABEL) trap. **E**, a simulated smFRET trace in the ABEL trap with fluorescence intensities for FRET donor in green and FRET acceptor in red shows one photon burst (start and end are indicated by grey bars). The first and last FRET level within the burst are omitted. Five full rotations are identified, each by a pair of medium FRET (L) plus high FRET ($H_1$, $H_2$) states in the proximity factor trace (P, blue). The upper trace shows the given FRET states in the Monte Carlo simulation (orange trace) in arbitrary units (a.u.).

**Figure 2**. **A**, scheme of the confocal ABEL trap including a continuous-wave 532 nm laser, polarizing beam splitter (PBS), mirrors (M), lenses (L), apertures (A), variable neutral density mirror (ND), two electro-optical beam deflectors (EOD x, y) driven by fast high-voltage amplifiers (HV), wave plate (λ/2), dichroic mirrors (DM 1, 2), oil immersion objective 60x with n.a. 1.42, 300 μm pinhole (PH), band pass and long pass filters (F 1, 2), single photon counting avalanche photodiodes (APD 1, 2), time-correlated single photon counting card (TCSPC) in one computer, field-programmable gate array (FPGA) card in a second computer, and low-voltage amplifiers (LV) for each of the four Pt-electrodes. Details are specified in the Supporting Information. **B**, the 32-point knight tour pattern with a schematic laser focus (green, shown at position 32, diameter approximately to scale). The focus position changed stepwise after 6.25 μs and followed the sequence →1→2→3→…→32→1→. **C**, color image of the active 32-point laser pattern in the trap region of the PDMS-on-glass ABEL trap chip (green squared area). Transitions from the cross-like

1 μm flat channels to the 80 μm deep channels for the electrodes are visible. Black scale bar is 20 μm. **D**, section of the solution-filled part of the PDMS-on-glass chip. The flat trap region (1 μm in height) and the flat channels in x- and y-direction are in the center, the deep channels (80 μm in height) with Pt-electrodes (black dots) are at the periphery [54]. Symmetric voltages with opposite signs are applied to x- or y-encoded electrodes by FPGA and LV. **E**, 3D model of the PDMS-on-glass chip with inserted Pt-electrodes (adopted from [55] with permission).

**Figure 3**. **A-C**, Photon bursts of reconstituted $F_oF_1$-liposomes labeled with Cy3B (green intensity traces) and Alexa Fluor 647 (red intensity traces) and held in the ABEL trap in the presence of 1 mM AMPPNP. Black traces are summed intensities. Blue traces above are the calculated proximity factor P values. The time bin is 1 ms for the thin line traces and 5 ms for the bold traces. **D**, P distribution of the FRET-labeled enzymes. **E**, 2D histogram of mean total brightness (summed intensities from both FRET donor and acceptor signals) *versus* P. **F**, 2D histogram of P *versus* photon burst duration.

**Figure 4**. **A-E**, five photon bursts of reconstituted $F_oF_1$-liposomes labeled with Cy3B and Alexa Fluor 647 and held in the ABEL trap in the presence of 1 mM ATP. Photon counts from FRET donor in green, FRET acceptor in red, sum intensity in grey, and proximity factor P in blue. Time binning is 1 ms. P values before and after the photon bursts of the trapped enzymes were shaded for clarity.

**Figure 5**. Distributions of individual ATP hydrolysis rates for FRET-labeled $F_oF_1$-ATP synthases for five ATP concentrations. **A**, 1 mM ATP, **B**, 100 μM ATP, **C**, 40 μM ATP, **D**, 20 μM ATP, **E**, 5 μM ATP. **Left diagrams**, time-dependence of individual catalytic rates; **middle diagrams**, histogram of rates; **right diagrams**, relation of the counted number of rotations during the FRET fluctuation period and the associated ATP hydrolysis rates.

**Figure 6**. Photon bursts of reconstituted $F_oF_1$-ATP synthases in the presence of 100 µM ATP and either 50 µM protonophore CCCP (**A**) or uncouplers 1 µM valinomycin and 1 µM nigericin (**C**). **B**, time-dependent distribution of individual ATP hydrolysis rates with CCCP. **D**, time-dependent distribution of individual ATP hydrolysis rates with valinomycin plus nigericin. **E**, box-and-whisker plot of all individual ATP hydrolysis rates at different ATP concentrations and with CCCP or valinomycin plus nigericin. **F**, Hanes-Woolf plot for ATP-dependent hydrolysis rates using the median values from figure **E**; error bars using $1^{st}$ and $3^{rd}$ quartiles (black triangles) from figure **E**. Expanded view for 5 µM to 100 µM ATP inserted.

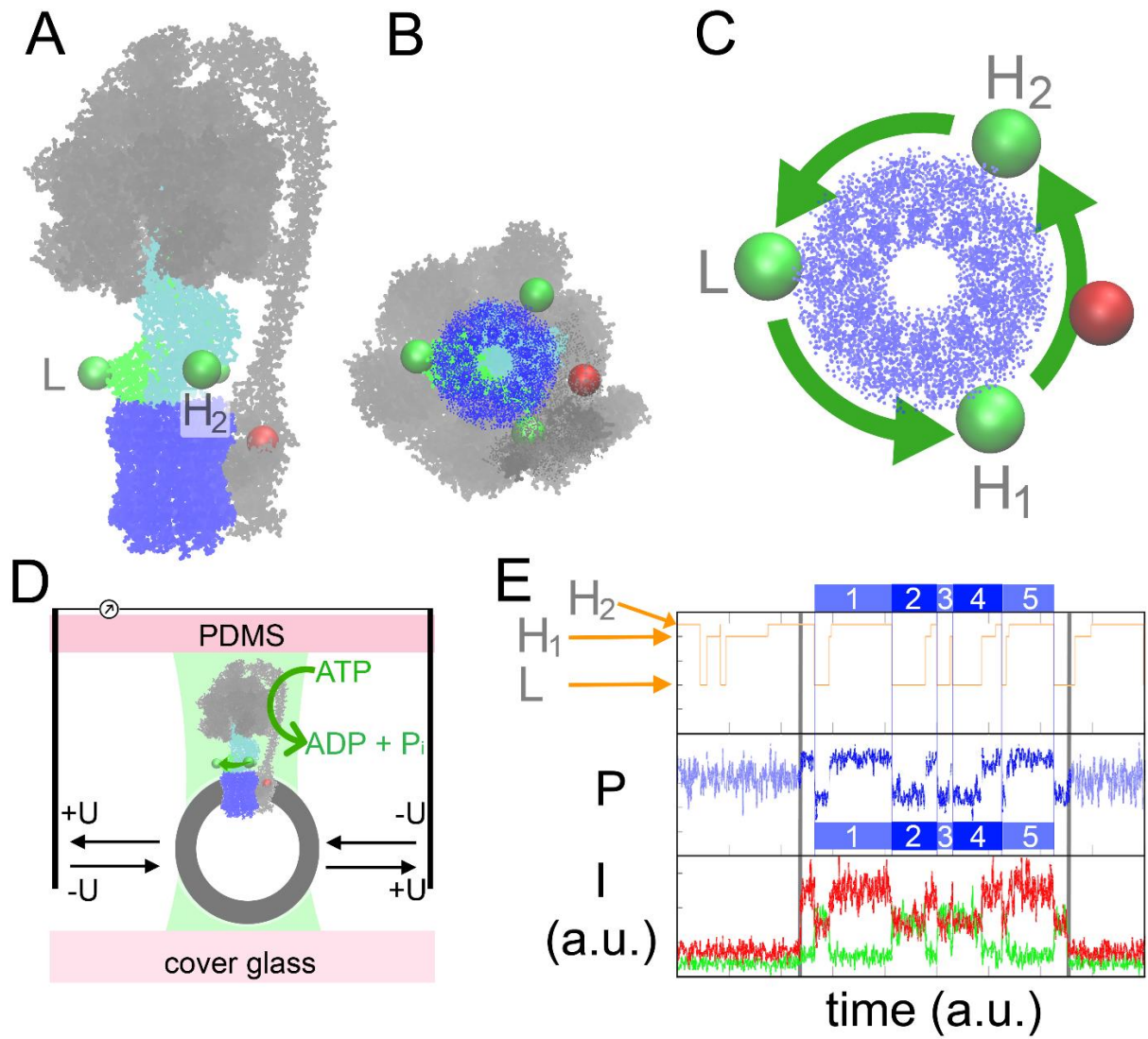

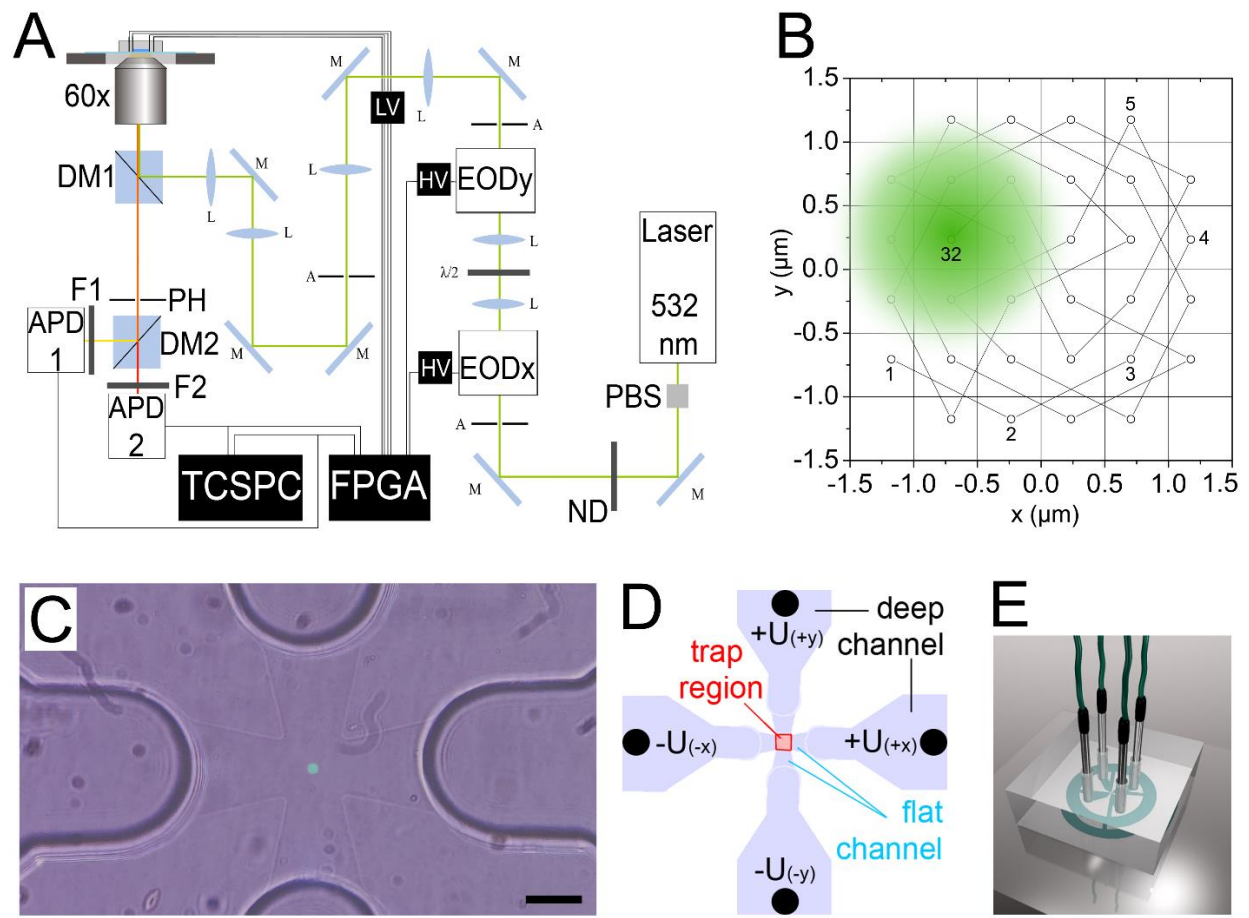

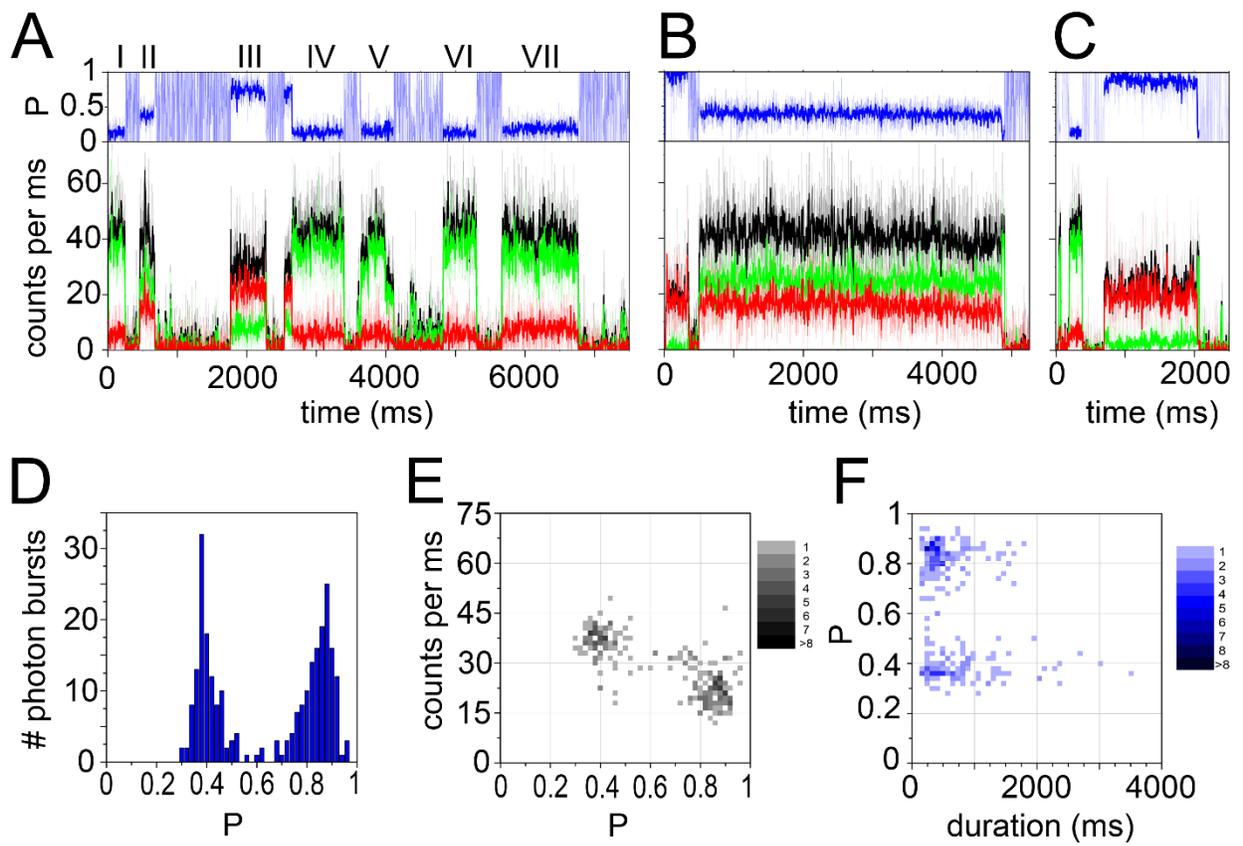

Figure 4

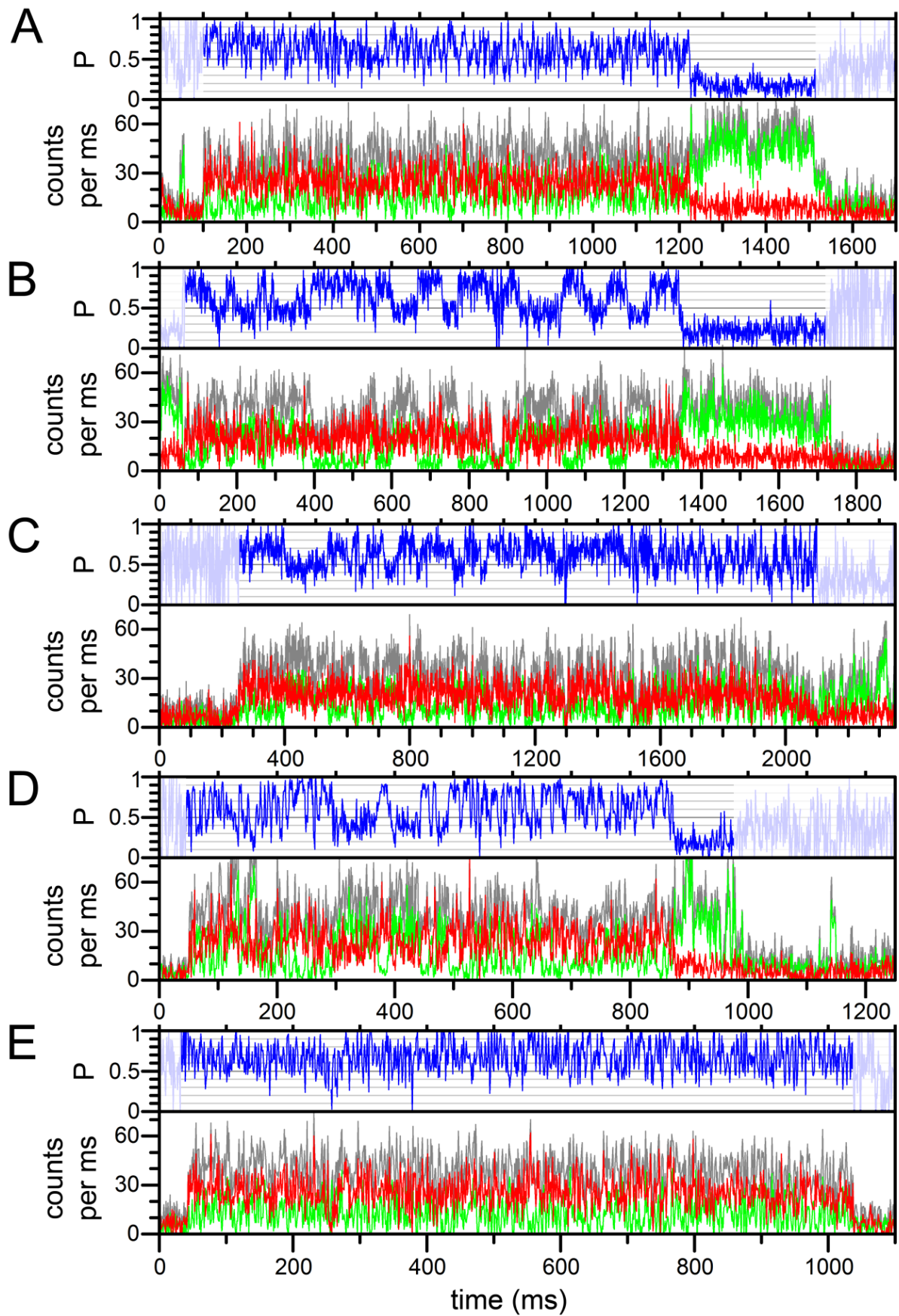

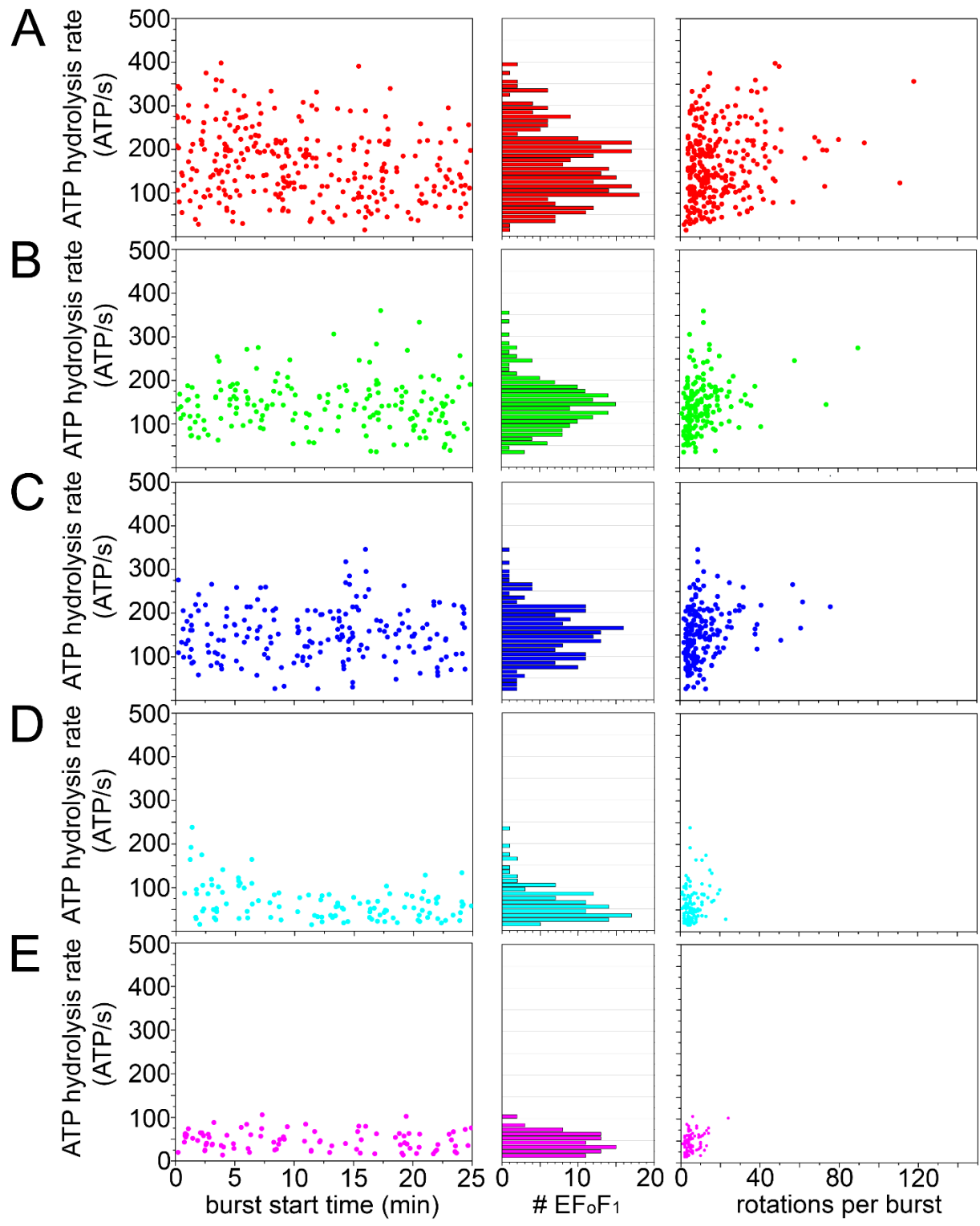

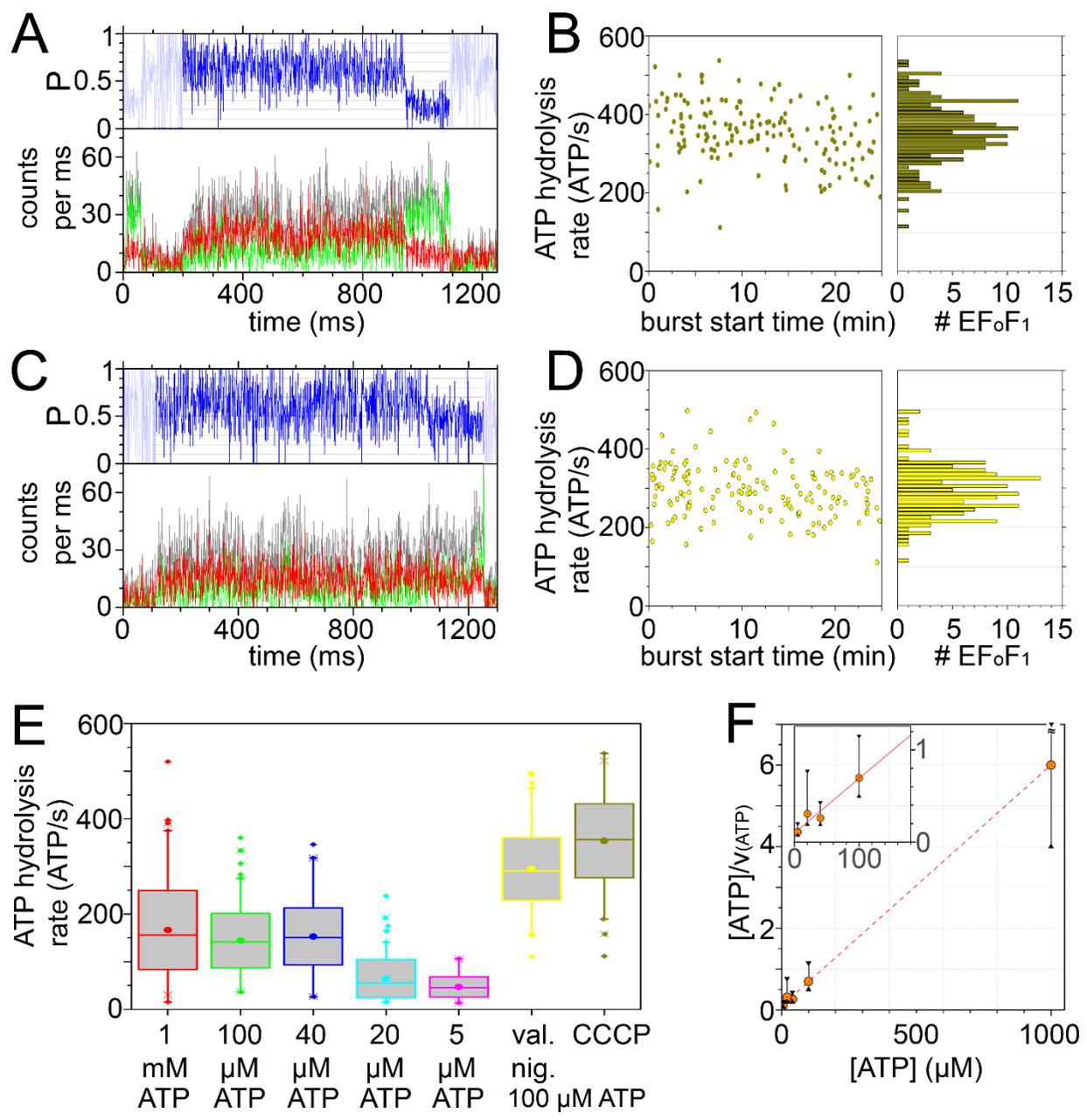

## 5. ACKNOWLEDGMENT

The authors thank A. E. Cohen and W. E. Moerner for liberally providing detailed design and laboratory protocols, the complete measurement software and multiple hardware advice to setup an ABEL trap. We thank all members of the single-molecule microscopy group who participated in biochemistry, and especially thank T. Rendler, N. Zarrabi, B. Su, M. Dienerowitz and A. Dathe who assembled different versions of the ABEL trap. We thank T. M. Duncan (SUNY Upstate Medical University) for valuable discussions. The ABEL trap was realized by funds from the Deutsche Forschungsgemeinschaft DFG (grants BO1891/10-2, BO1891/15-1, BO1891/16-1, BO1891/18-2 to M.B.) and was supported by an ACP Explore project (M.B. together with J. Limpert) within the ProExcellence initiative ACP2020 from the State of Thuringia.

## 6. ASSOCIATED CONTENT

**Supporting Information**. An associated PDF file with 16 supplemental figures is available free of charge online, comprising preparation of FRET-labeled $F_oF_1$-ATP synthase from *E. coli* including fluorophore labeling, reconstitution, catalytic activity measurements of ATP hydrolysis and synthesis, the confocal ABEL trap setup, PDMS-on-glass sample chamber preparation, photophysics of FRET fluorophores bound to the enzyme with and without the antioxidant *trolox*, photon bursts of single FRET-labeled $F_oF_1$-ATP synthases in the presence of 100 µM, 40 µM, 20 µM, or 5 µM ATP, distributions of fluctuating FRET level durations, and Lineweaver-Burk analysis of smFRET-based ATP hydrolysis rates.

# 7. AUTHOR INFORMATION

**Corresponding Author**

Corresponding author: Michael Börsch. Single-Molecule Microscopy Group, Jena University Hospital, Nonnenplan 2 – 4, 07743 Jena, Germany.

Phone: +49 3641 9396618; Fax: +49 3641 9396621; Email: michael.boersch@med.uni-jena.de

**Author Contributions**

The manuscript was written through contributions of both authors. Both authors have given approval to the final version of the manuscript.

**Notes**

The authors declare no competing financial interest.

# 8. ABBREVIATIONS

ATP, adenosine triphosphate; εCTD, Carboxy-terminal domain of the ε-subunit of $F_oF_1$-ATP synthase; a-CT, carboxy-terminus of the *a*-subunit of $F_oF_1$-ATP synthase; *pmf*, proton motive force; CCCP, carbonyl cyanide m-chlorophenyl hydrazone; PDMS, polydimethylsiloxane; ABEL trap, anti-Brownian electrokinetic trap; smFRET, single-molecule Förster resonance energy transfer; P, proximity factor; TMR, tetramethylrhodamine; Cy3B, cyanine 3B; FWHM, full width at half maximum; TCSPC, time-correlated single photon counting; FPGA, field-programmable gate array.

## 10. TOC GRAPHIC

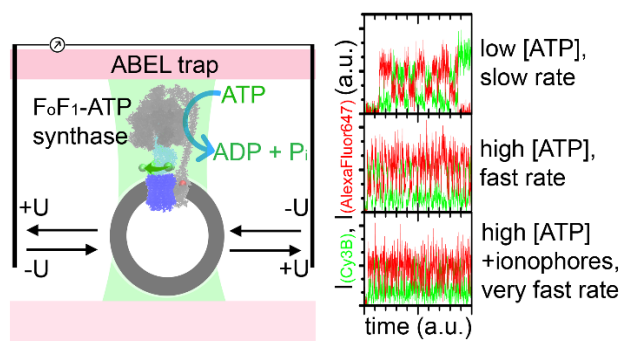

# Supporting Information

# Fast ATP-dependent Subunit Rotation in Reconstituted $F_oF_1$-ATP Synthase Trapped in Solution

*Thomas Heitkamp, Michael Börsch*

Single-Molecule Microscopy Group, Jena University Hospital, 07743 Jena, Germany.

Email: michael.boersch@med.uni-jena.de

Phone +49 3641 9396618, Fax +49 3641 9396621

**Content**

(1) Preparation of FRET-labeled $F_oF_1$-ATP synthase from *E. coli* including fluorophore labeling, reconstitution, catalytic activity measurements of ATP hydrolysis and synthesis.

(2) Confocal ABEL trap setup, PDMS-on-glass sample chamber preparation, photophysics of FRET fluorophores bound to the enzyme.

(3) Photophysics in the presence of the antioxidant *trolox*.

(4) Photon bursts of single FRET-labeled $F_oF_1$-ATP synthases in the presence of 100 µM, 40 µM, 20 µM, or 5 µM ATP, distributions of fluctuating FRET level durations, Lineweaver-Burk analysis of smFRET-based ATP hydrolysis rates.

# 1. Preparation of FRET-labeled $F_oF_1$-ATP synthase reconstituted in liposomes

**1.1 Plasmids and *E. coli* expression strain.** To achieve specific labeling at defined cysteine positions within the $F_o$ complex and the $F_1$ complex of *E. coli* $F_oF_1$-ATP synthase (or $EF_oF_1$, respectively), the two complexes had to be purified and labeled separately. This approach was developed previously [1]. Here, after reconstituting $EF_oF_1$-ATP synthase labeled at the a-subunit into liposomes, its unlabeled $F_1$ complex was stripped and replaced by the $F_1$ complex labeled at the ε-subunit to yield the FRET-labeled reconstituted enzyme. $F_oF_1$-ATP synthase carrying a cysteine at the extended C-terminus of the a-subunit (a-GAACA; $F_o$-(a-SH)$F_1$-ATP synthase in the following) and the $F_1$ complexes carrying the H56C substitution in the ε-subunit were produced using the corresponding plasmids pMB14 (a-GAACA) [2] and pRAP100 (εH56C) [3]. Plasmids pMB14 and pRAP100 are derivatives of pRA100 [4]. Plasmids were transformed into *Escherichia coli* strain RA1 (*F⁻ thi rpsL gal Δ(cyoABCDE) 456::KAN Δ(atpB-atpC)ilv:Tn10*) [5], and cells were grown in a 10L fermenter (FerMac 320, Electrolab Biotech, UK) [6].

**1.2. Purification of $F_o$-(a-SH)$F_1$-ATP synthase.** Tag-free $F_o$-(a-SH)$F_1$-ATP synthase was purified using three chromatography steps [6]. In brief, $F_oF_1$-containing membranes were isolated, washed to remove membrane-associated proteins and solubilized with n-dodecyl β-D-maltoside (DDM) as the detergent (SDS-PAGE shown in Figure S1 below, lane "Mem"). Ultracentrifugation was applied to remove insoluble material. Solubilized membrane proteins (Figure S1, lane "Sol") were concentrated by ammonium sulfate precipitation and subsequently separated from lipids, nucleotides and salts by size exclusion chromatography (SEC) using a self-packed XK26/40 Sephacryl S-300 column (GE Healthcare, USA). The eluted protein fractions (Figure S1, lane "1. SEC") were loaded separately onto an anion exchange (IEX) column Poros HQ 20 (4.6 x 100 mm, Applied Biosystems, USA). The $F_oF_1$-containing peak fractions were pooled (Figure S1, lane "IEX"), concentrated by ammonium sulfate precipitation and subsequently loaded onto a Tricorn Superose 6 10/300 GL size exclusion column (GE Healthcare, USA) for refinement. The second SEC was mainly used for desalting and for separation of dissociated $F_o$ and $F_1$ fractions. The protein fractions containing intact, pure $F_oF_1$ with all subunits (Figure S1, lane "2. SEC") were separately shock-frozen in liquid nitrogen and stored at -80°C for subsequent labeling.

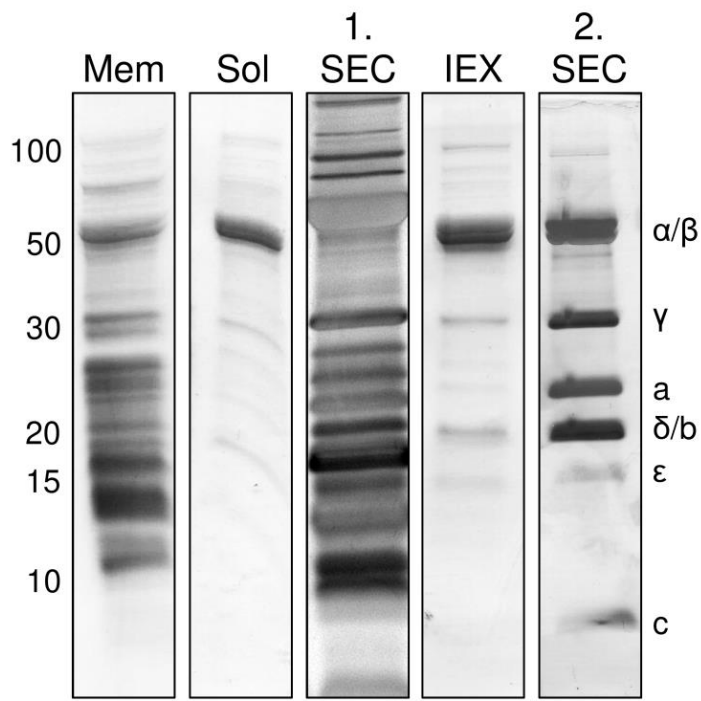

**Figure S1.** SDS-PAGE of the $F_o$-(a-SH)$F_1$-ATP synthase purification steps from *E. coli* membranes. Acrylamide concentration was 12 %, staining with coomassie-R250. Molecular weight numbers in kDa on the left, $F_oF_1$ subunit assignment on the right.

**1.3. Purification of $F_1$-($\varepsilon$H56C) complexes.** Tag-free $F_1$-($\varepsilon$H56C) complexes were purified using established protocols [3, 7] with minor modifications. Briefly, $F_oF_1$-containing membranes were isolated and washed with low ionic strength buffers (SDS-PAGE shown in Figure S2 below, lane "Mem"). The soluble $F_1$ complex was then stripped from the membrane-embedded $F_o$ complex using buffer without $MgCl_2$ and 4-aminobenzamidine. The soluble fraction (Figure S2, lane "Strip") was loaded on a self-packed Poros HQ 20 (10 x 200 mm) anion exchange column (Applied Biosystems, USA). The $F_1$ complexes were almost pure (Figure S2, lane "IEX"). Only two impurities with molecular weights between 12 and 20 kDa and some faint protein bands in the high molecular weight range were remaining. Therefore, $F_1$ complexes were concentrated by ammonium sulfate precipitation and loaded on a Tricorn Superdex 200 10/300 GL column (GE Healthcare, USA) for size exclusion chromatography. The major protein peak fractions containing pure $F_1$ complexes (Figure S2, lane "SEC") were pooled, concentrated using an Amicon Ultra-15 30K (Merck Millipore, USA), shock-frozen in liquid nitrogen and stored at -80 °C.

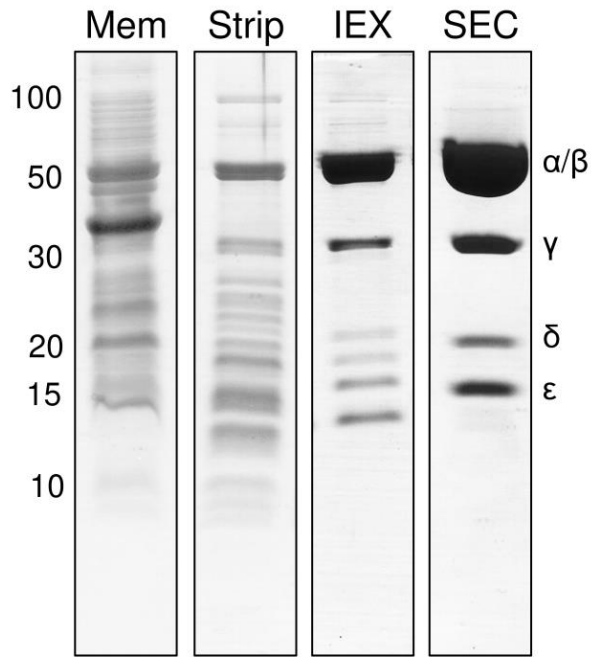

**Figure S2.** SDS-PAGE of $F_1$-(εH56C) complex purification steps from *E. coli* membranes. Acrylamide concentration was 12 %, staining with coomassie-R250. Molecular weight numbers in kDa on the left, $F_1$ subunit assignment on the right.

**1.4. Fluorescence labeling of the a- and ε-subunit.** Cysteine labeling reactions were carried out as described [3, 6]. Purified $F_o$-(a-SH)$F_1$-ATP synthase or purified $F_1$-(εH56C) were first concentrated by precipitation with $(NH_4)_2SO_4$, dissolved in 75 µl 50 mM MOPS-NaOH pH 7.0, with 0.1 mM $MgCl_2$ and, in the case of $F_oF_1$, with 0.1% (w/v) DDM ("labeling buffer with DDM"). The resolved proteins were desalted by size exclusion chromatography using 1 ml columns filled with Sephadex G50 medium (GE Healthcare, USA), equilibrated by gravity flow with labeling buffer and pre-spinned for 2 min with 250 × g at room temperature. The resolved proteins were eluted by centrifugation for 2.5 min at 445 × g. Concentrations of the desalted enzymes were calculated by absorption measurements at 280 nm (Lambda 650, PerkinElmer, USA) using a molar extinction coefficient ($ε_{280nm}$) of 303,130 $L·mol^{-1}·cm^{-1}$ for $F_oF_1$ and 206,060 $L·mol^{-1}·cm^{-1}$ for $F_1$.

The cysteine in the a-subunit of $F_oF_1$ was labeled with Alexa Fluor 647 $C_2$-maleimide (Thermo Fischer Scientific Inc., USA) and the cysteine in the ε-subunit of $F_1$ was labeled with Cy3B maleimide (GE Healthcare, USA). Labeling was carried out in labeling buffer with a five-fold molar excess of Tris(2-carboxyethyl)phosphine (TCEP) to protein on ice, in a total volume of 100 µl using a protein concentration of 10 µM $F_oF_1$ or 14 µM $F_1$ and a 1:1.1 stoichiometric ratio of enzyme to dye. Reactions were stopped after 15 minutes by three consecutive Sephadex G50 medium spin columns. Labeling efficiencies were determined by comparing absorption at 280 nm and dye absorption at its maximum (Cy3B: $\varepsilon_{559nm}$ = 130,000 L·mol$^{-1}$·cm$^{-1}$; Alexa Fluor 647: $\varepsilon_{654nm}$ = 265,000 L·mol$^{-1}$·cm$^{-1}$). Shown in Figure S3, labeling the ε-subunit with Cy3B-maleimide was achieved with an efficiency of 73.7%. Labeling efficiency of the a-subunit with Alexa Fluor 647-maleimide was 31.9%. We noticed that the spectrum of enzyme-bound Alexa Fluor 647 was shifted about 7 nm towards longer wavelengths.

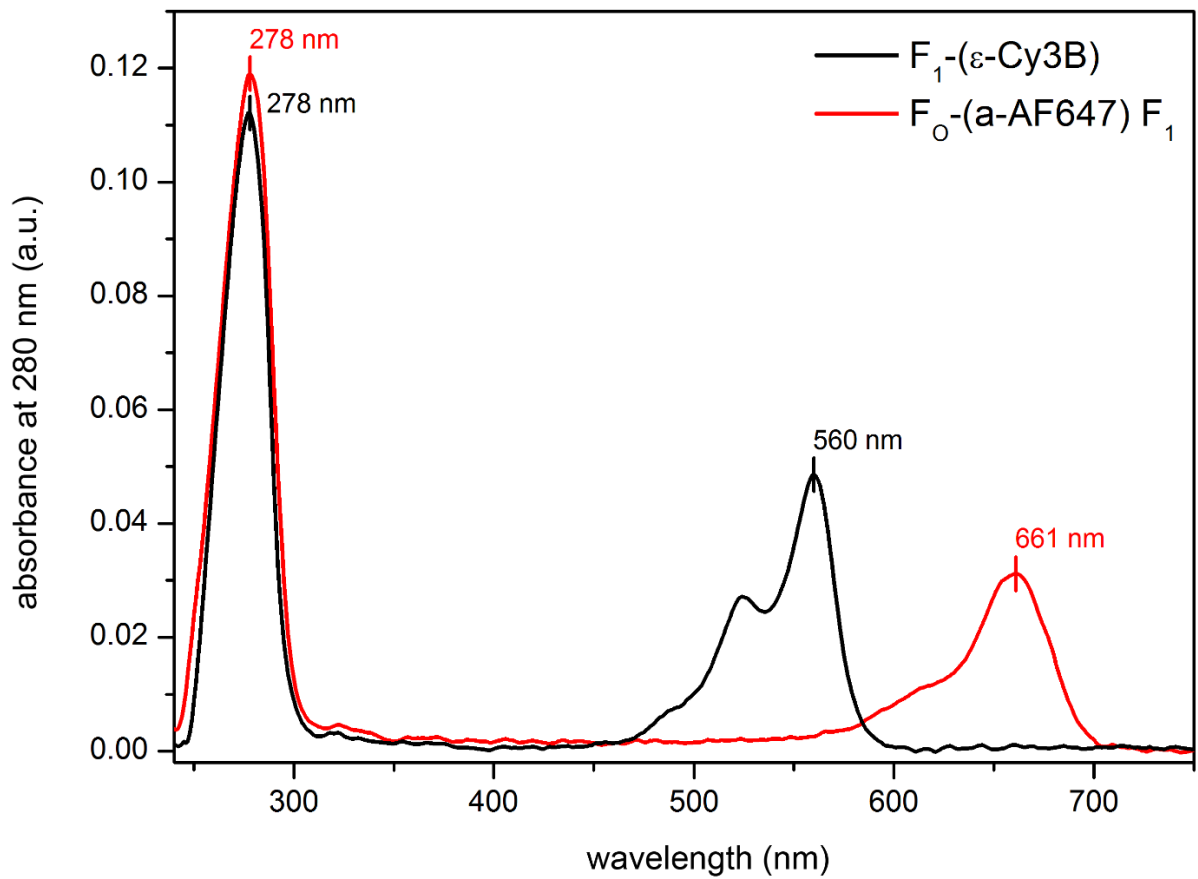

**Figure S3.** Absorbance spectra of Cy3B-labeled $F_1$-(εH56C) complexes (black curve) and of Alexa Fluor 647-labeled $F_o$-(a-SH)$F_1$-ATP synthases (red curve).

Labeling specificity was checked by SDS-PAGE with subsequent fluorography (Figure S4). Accordingly, both labeling approaches were highly specific. The labeled proteins were adjusted with 10% (v/v) glycerol in buffer, flash frozen as 5 µl aliquots in liquid $N_2$ and stored at -80°C.

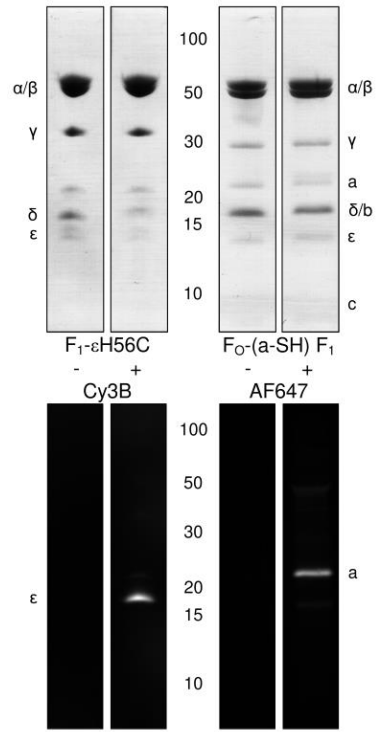

**Figure S4.** SDS-PAGE of $F_1$-($\varepsilon$H56C) complexes and of $F_o$-(a-SH)$F_1$-ATP synthase before (-) and after (+) labeling with Cy3B or Alexa Fluor 647, respectively. Upper panels are white-light transmission images of gels stained with coomassie-R250. Lower panels are fluorescence images showing the specificity of the labeling.

## 1.5. Reconstitution of $F_oF_1$-ATP synthase and exchange of $F_1$ complexes to yield FRET-labeled $F_oF_1$-ATP synthase in liposomes.

First, Alexa Fluor 647-labeled $EF_oF_1$ were reconstituted into pre-formed liposomes [6]. Liposomes were generated by evaporating chloroform solutions of phosphatidylcholine (Lipoid GmbH, Germany) and phosphatidic acid (Sigma-Aldrich, Germany) in a mass ratio of 19:1. The dried lipid film was resolved to 18 g/l in 10 mM Tricine-NaOH pH 8.0, 0.1 mM EDTA, 0.5 mM DTT, 7.2 g/l cholic acid and 3.6 g/l desoxycholate. The suspension was sonicated and dialyzed at 30°C for 5 h against the 4,000-fold volume of 10 mM Tricine-NaOH pH 8.0, 0.2 mM EDTA, 0.25 mM DTT and 2.5 mM $MgCl_2$. The reconstitution efficiency is highly dependent on the reconstitution volumes and the sizes of the tubes used. To obtain a final volume of 1.2 ml proteoliposomes, three 2 ml Eppendorf tubes filled with 400 µl reconstitution assay were prepared in parallel. In each Eppendorf tube, 200 µl of the pre-formed liposomes, 1 µl of 1 M $MgCl_2$ (2.5 mM final concentration), "reconstitution buffer" (20 mM Tricine-NaOH pH 8.0, 20 mM succinate, 50 mM NaCl and 0.6 mM KCl) to fill up to 400 µl end volume and 20 nM labeled $EF_oF_1$ were mixed in exactly this order. Then, 32 µl of 10% (v/v) Triton-X-100 (i.e., 0.8% (v/v) final concentration) were added under vigorous shaking (Vortex-Genie 2, Scientific Industries,

USA). After 1 h incubation with gentle shaking, 128 mg of pre-treated BioBeads SM-2 (Biorad, USA) were added to each Eppendorf tube to remove the detergent. After 1 hour on BioBeads, the proteoliposomes were separated from the BioBeads and pooled.

Second, the unlabeled $F_1$ complex was removed from reconstituted $EF_oF_1$ as described [8]. Proteoliposomes were diluted 25-fold with a low ionic strength buffer (1 mM Tricine–NaOH pH 8.0, 1 mM DTT, 0.5 mM EDTA, and 4% (v/v) glycerol) without $MgCl_2$, incubated for 1 h at room temperature and centrifuged at $300,000 \times g$ for 1.5 h at room temperature. Following gentle resuspension, this procedure was repeated for another two times. The remaining $F_o$-liposomes were resuspended in the buffer above with 10% glycerol to yield a $F_o$ concentration of 125-150 nM and stored overnight in the dark.

Third, rebinding of Cy3B-labeled $F_1$ to $F_o$-liposomes was achieved as published [8]. The Alexa Fluor 647-$F_o$-liposome suspension was incubated with Cy3B-labeled $F_1$ in solution at a molar excess of three $F_1$ per $F_o$ in the presence of 2.5 mM $MgCl_2$ and 50 mM NaCl for 45 min at 37°C, followed by 90 min incubation at 0°C. Unbound $F_1$ was removed by two ultracentrifugation runs (90 min, $300,000 \times g$, 4°C). Each time the pelleted proteoliposomes were resuspended in 20 mM Tricine-NaOH (pH 8.0), 20 mM succinic acid, 50 mM NaCl, 0.6 mM KCl, 2.5 mM $MgCl_2$ and 4% (v/v) glycerol. The proteoliposomes were adjusted to 10% (v/v) glycerol, flash-frozen as 5 µl aliquots in liquid nitrogen and stored at −80°C. The final concentration of FRET-labeled $F_oF_1$-liposomes was estimated to ~40 nM.

**1.6. Measurement of ATP hydrolysis rates**. ATP hydrolysis was measured using a continuous coupled enzymatic assay [6, 9, 10]. Solubilized enzyme (3-10 nM) was added to the 37°C pre-warmed reaction buffer (100 mM Tris-HCl pH 8.0, 25 mM KCl, 4 mM $MgCl_2$, 2.5 mM phosphoenolpyruvate, 18 units/ml pyruvate kinase, 16 units/ml lactate dehydrogenase, 2 mM ATP and 0.4 mM NADH), and ATP hydrolysis was followed by the decrease of NADH absorbance. The ATP hydrolysis rate was calculated from the linear slope using an extinction coefficient of $6,220$ L $mol^{-1}$ $cm^{-1}$ for NADH. By adding N,N-dimethyl-n-dodecylamine N-oxide (LDAO) to a final concentration of 0.6% (v/v), possible activation of the enzyme as an indicator for purification quality was verified [11, 12]. Solubilized $F_oF_1$ in detergent showed significantly lower ATP hydrolysis rates than $F_1$ (Figure S5). However, LDAO-induced activation of $F_oF_1$ was in good agreement with reported values. Labeling with fluorescent dyes had no significant effect on ATP hydrolysis rates of the enzymes.

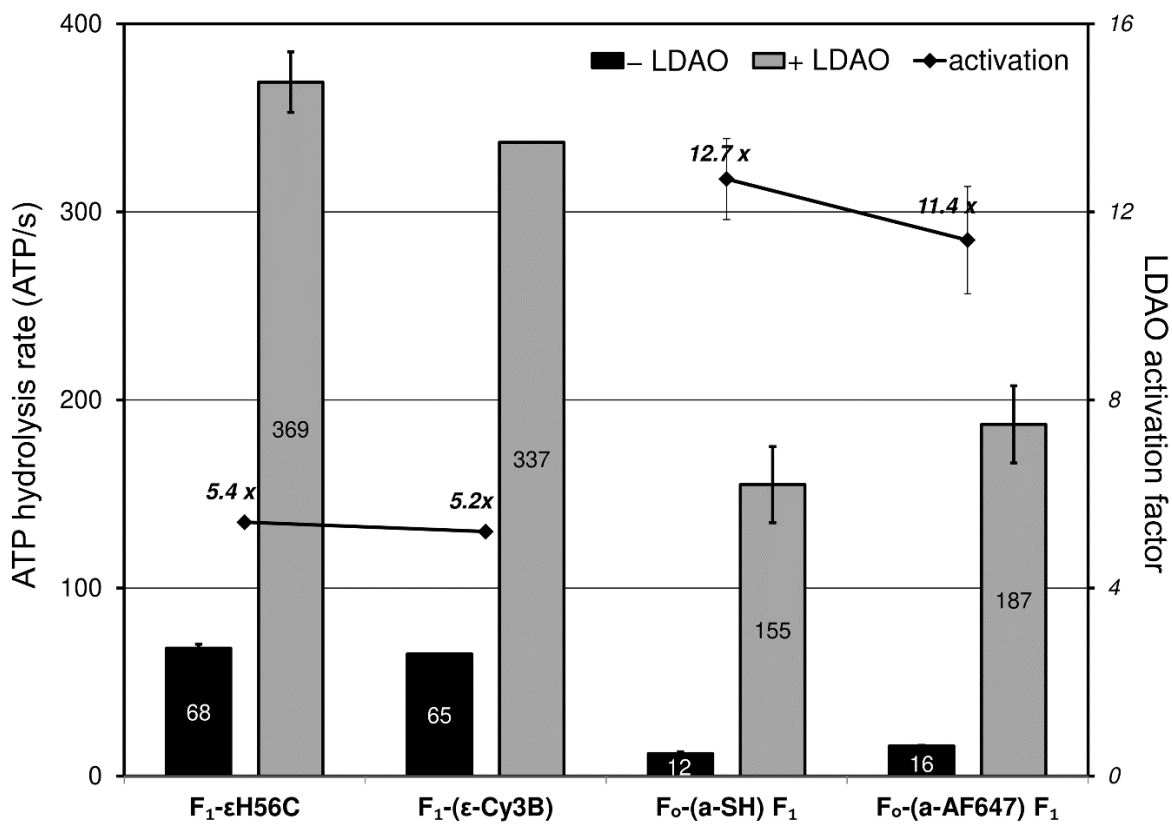

**Figure S5.** ATP hydrolysis rates of $F_1$-($\varepsilon$H56C) complexes and DDM-solubilized $F_o$-(a-SH)$F_1$-ATP synthase before and after fluorescence labeling (black bars). Activation of ATP hydrolysis rates in the presence of LDAO are shown as grey bars, with associated activation factors shown as black diamonds.

**1.7. Measurement of ATP synthesis rates.** ATP synthesis rates were determined as described [13] using the luciferin/luciferase luminescence assay. The cysteine mutation at the C-terminus of the a-subunit and fluorophore labeling did not affect ATP synthesis activity of reconstituted $F_oF_1$-ATP synthase compared to the "wildtype" enzyme (Figure S6). However, after rebinding $F_1$ to $F_o$-liposomes the FRET-labeled enzyme showed less than half of the ATP synthesis activity. This was likely a $F_oF_1$ concentration artefact due to the loss of proteoliposomes during each of the ultracentrifugation steps and had been discussed previously [3, 10]. Therefore, a mean ATP synthesis rate 16 ±2 ATP·s$^{-1}$ for the reassembled FRET-labeled enzyme is in good agreement either with previously reported 21 ±4 ATP·s$^{-1}$ for a similar FRET-labeled $F_oF_1$-ATP synthase comprising the same cysteine mutation $\varepsilon$H56C in $F_1$, cysteine mutation in the b-subunits, and rebinding of $F_1$ to $F_o$ procedures [3], or with a ATP synthesis rate of 32 ±7 ATP·s$^{-1}$ for $F_oF_1$ with EGFP (or mNeonGreen, respectively) fused to the C-terminus of the a-subunit but without rebinding $F_1$ [6, 14].

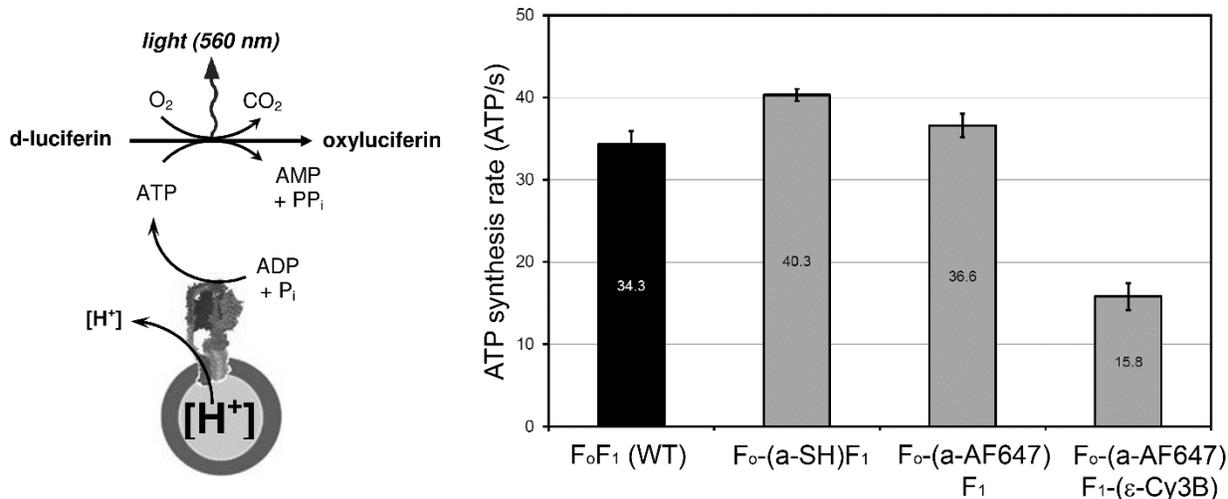

**Figure S6.** Left, principle of the luciferin/luciferase assay for ATP synthesis measurements. Right, comparison of ATP synthesis rates of "wild-type" $F_oF_1$ (black bar), of $F_o$-(a-SH)$F_1$-ATP synthase before and after labeling with Alexa Fluor 647 and of the reassembled FRET-labeled $F_oF_1$-ATP synthase (grey bars).

## 2. The confocal ABEL trap setup

**2.1. Confocal setup**. We have built different ABEL traps with APD detectors using a variety of lasers and associated optical filter sets [15, 16, 17]. Here, the beam diameter of a continuous-wave 532 nm laser (Compass 315M, Coherent) was collimated and adjusted to less than 1 mm by a two-lens telescope and linearly polarized by a polarizing beam splitter cube [18]. Laser power was set to 40 µW using a variable metallic neutral density filter. Beam steering was achieved by two consecutive electro-optical deflectors (M310A, Conoptics) with an achromatic $\lambda/2$ wave plate in between. Each EOD was powered by a fast high-voltage amplifier (7602 M, Krohn-Hite) and controlled by the field-programmable gate array (FPGA card PCIe-7852R, National Instruments). Our confocal ABEL trap setup is shown in Figure S7. The FPGA LabView software to run the ABEL trap [19] was used with minor modifications to implement the 32-point knight pattern in the focus plane [20] and to extend the number of trapping detector channels up to four.

Entering the inverted microscope (IX 71, Olympus) *via* the epifluorescence port, the laser beam was directed to a 60x oil immersion objective with n.a. 1.42 (PlanAPO, Olympus) using a dichroic mirror (z 532 RD, F43-537, AHF). From the focal plane, fluorescence was focused to a 300 µm pinhole by the tube lens on the left side port of the IX71 and separated into two spectral channels by a dichroic beam splitter at 640 nm (zt 640 RDC, F48-640, AHF). FRET donor Cy3B photons were detected in the range between 545 nm and 620 nm (582/75 BrightLine HC, F37-582, AHF) and FRET acceptor Alexa Fluor 647 photons for wavelengths λ > 647 nm (Raman RazorEdge LP 647 RU, F76-647, AHF). Given filter specifications and available fluorescence spectra at *www.FPbase.org*, the filter-based spectral detection efficiency for Cy3B was estimated to 69.2% in the FRET donor channel plus a spectral crosstalk of 11.4% into the acceptor channel. Alexa Fluor 647 was detected in the acceptor channel with 92.3% efficiency and without a crosstalk to the donor channel. Time-correlated single photon counting (TCSPC) by two avalanche photodiode detectors (SPCM AQRH-14, Excelitas) was achieved with 164 ps time resolution using the TCSPC inputs of the counter electronics (DPC-230, Becker&Hickl) and an external sync signal. Both APDs were mounted on a single 3D-adjustable mechanical stage system (home-built with components from OWIS) that also contained the pinhole. Positioning of the ABEL trap chip on the microscope was possible using a mechanical stage (x, y translation with μm resolution) and the z-positioner of the IX71, plus an 3D piezo scanner (P-527.3CD, Physik Instrumente) with sub-nm resolution mounted on top of the mechanical stage.

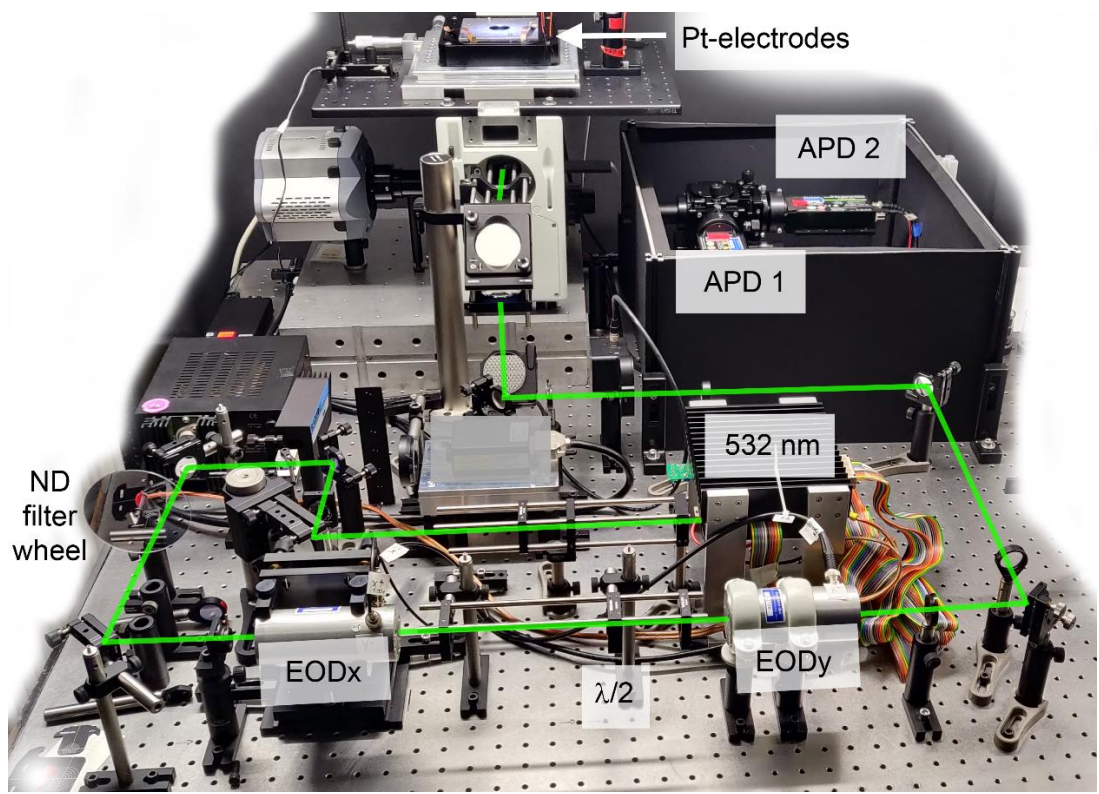

**Figure S7.** Photographic image of the confocal ABEL trap setup. Laser beam path in green.

**2.2. FPGA-based ABEL trap.** The multiplexed APD signals were counted in parallel by the FPGA card in a separate measurement computer. A Kalman filter in the FPGA software correlated the APD signals with respect to the actual laser focus in the knight tour pattern. The FPGA software estimated the position of the fluorescent particle and generated the feedback voltages to the four Pt-electrodes. The low-voltage amplifier limit was set to $\pm 10$V at the electrodes. This version of the FPGA LabView software required optimization of several parameter settings to trap the negatively-charged proteoliposomes for long times. For the 32-point knight tour pattern in a 6 x 6 array, the distance between the focus positions was set to 0.47 μm, the repetition rate of the full pattern was set to either 5 or 7 kHz, the dimensions of the pattern in the focal plane were 2.34 x 2.34 μm, the waist of the laser focus was estimated as 0.6 μm using 100-nm-tetraspeck bead scanning of immobilized beads, and the EOD deflection scale was measured as 0.117 μm/V. The FPGA has a 0 to 10 V output, i.e., 10 V correspond to 10V x 0.117 μm/V = 1.17 μm deflection of the laser focus in the focal plane. The deflection scale is given from the center of the laser pattern into either +x or -x (or +y or -y) direction. Therefore the given EOD deflection scale describes the dimensions of the knight tour pattern, i.e., here we covered 2.34 x 2.34 μm with 32 focus points. The values for the estimated diffusion constant in the software in order to trap proteoliposomes was set to 5 μm$^2$/s, and the estimated electric mobility was set to -80 μm/s/V.

Trapping of FRET-labeled $F_oF_1$-ATP synthases was achieved using the combined photon counts from donor and acceptor channel. Qualitatively, "good trapping" of proteoliposomes was classified by online monitoring of three indicators simultaneously as shown in three panels of the graphical user interface: (1) maximum brightness (i.e., combined donor and acceptor photon count rates) and minimal intensity fluctuations of a trapped particle, (2) the generated feedback voltages (symmetric for ±x and  ±y voltages, low voltages fluctuating around 0 V during trapping in solution), and (3) the visualized trapping position within the center (0,0 position) of knight tour pattern. The quality of the PDMS-on-glass sample chamber (i.e., the symmetry of the thin channels in the cross-like trapping region, see below) and the embedding of the electrodes in the sample buffer could be assessed. Furthermore, proteoliposomes which were temporarily attached to the surface, could be revealed in real time. Brightness loss in combination with shorter trapping durations indicated a z-drift of the sample chamber with respect to the focal plane, and prompted manual adjustment of the z-position by quick refocusing using the ocular port of the microscope.

**2.3. PDMS-on-glass chips.** The custom-made wafer used to mold the PDMS part of the sample chamber was produced at the Leibniz Institute for Photonic Technologies (IPHT) Jena, FAG 53 microsystem technologies. The current wafer design followed previously optimized structures [21, 22, 23]. Here, a 100-mm round floated borosilicate glass wafer (BOROFLOAT®, Schott) with 1.1 mm thickness comprising 26 chips with 14.15 x 14.15 mm size was employed. Each numbered chip consisted of two photoresist coatings, i.e. a 1.1 μm thin layer for the cross-like trapping region and an 82 μm thick layer for the deep channels to the electrodes and the channels to the outer circle to compensate hydrodynamic flow. A color image of the wafer with PDMS on top is shown in Figure S8 A.

A PDMS mixture with a 10:1 ratio of polymer:hardener (Sylgard 184 elastomer, Dow Corning) was poured over the wafer in a petri dish and cured for 4 h at 70°C. The PDMS layer was carefully removed from the wafer, and each chip was cut out using a scalpel. Holes for the electrodes were punched using a sharpened needle. A single PDMS chip was sonicated in pure acetone, rinsed with deionized water and dried in a nitrogen stream. A 32 x 24 mm cover glass with defined thickness (#1.5) was rinsed with deionized water and dried in a nitrogen stream. PDMS and cover glass were evacuated for 15 to 20 min and were plasma-etched for 90 seconds. Because we studied time-dependent single-molecule enzyme kinetics, each sample chamber was prepared immediately before starting the measurements on the ABEL trap. Briefly, after PDMS was bonded to the cover glass, the sample chamber was filled quickly with 10 μl of the proteoliposome solution using a thin gel-loader tip, the sample chamber was clamped on the microscope, Pt-electrodes were put into the four holes, the laser focus pattern was positioned accurately in the center of the flat trapping region and the z-position of the focus was aligned. All procedures were completed within 2 minutes. Then smFRET data were recorded consecutively in five 5-min runs to limit the TCSPC file size. A color image of the PDMS-on-glass sample chamber with electrodes clamped on top of the microscope is shown in Figure S8 B.

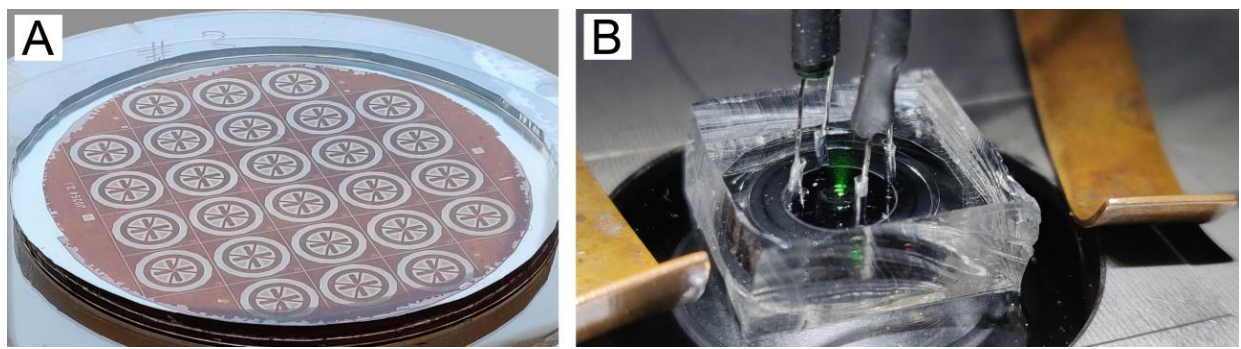

**Figure S8. A,** wafer with PDMS layer for 26 sample chambers. **B**, mounted ABEL trap chip with inserted electrodes and 532 nm laser on.

**2.4. 32-point knight tour pattern**. The focus pattern expanded the size of 532 nm excitation volume by a factor of ~5 as estimated from FCS measurements of Rhodamine 6 G (R6G) in aqueous solution (Figure S9 A). This was expected from the laser focus waist of 0.6 μm and the given pattern dimensions of 2.34 x 2.34 μm. Due to the thin laser beam diameter < 1 mm underfilling the back aperture of the objective, the mean diffusion time of R6G was $\tau_{(D)} = 0.35$ ms with the EODs turned off (black curve, FCS fitting shown as red curve). The FCS fitting model included a triplet term with $\tau_{(T)} = 3$ μs. The diffusion time is indicated by the position of the black cross in Figure S9 A. Turning the knight tour pattern on altered the autocorrelation function G($\tau$) of R6G significantly (blue curve, FCS fitting shown as red solid line plus approximation to short correlation times as dotted red curve). The mean diffusion time shifted to $\tau_{(D)} = 1.67$ ms (marked by the blue cross), and several strong correlations were visible. The 7 kHz repetition rate of the pattern corresponded to the peak at a correlation time t = 0.14 ms as marked by the blue asterisk. Additional faster correlations were revealed due to the 4.08 μs dwell for each point and a repeating overlap of the focus points within the knight tour pattern.

Similarly the diffusion times of proteoliposomes in the cross-like shallow trapping region of the sample chip were extended, i.e. from $\tau_{(D)} = 35$ ms with a single fixed laser position and inactive feedback (black curve in Figure S9 B) to $\tau_{(D)} \sim 270$ ms with the knight tour pattern and the electrode feedback for trapping activated. Due to a lower repetition rate of 5 kHz, the related correlation peak was found at a correlation time of 0.2 ms (purple asterisk).

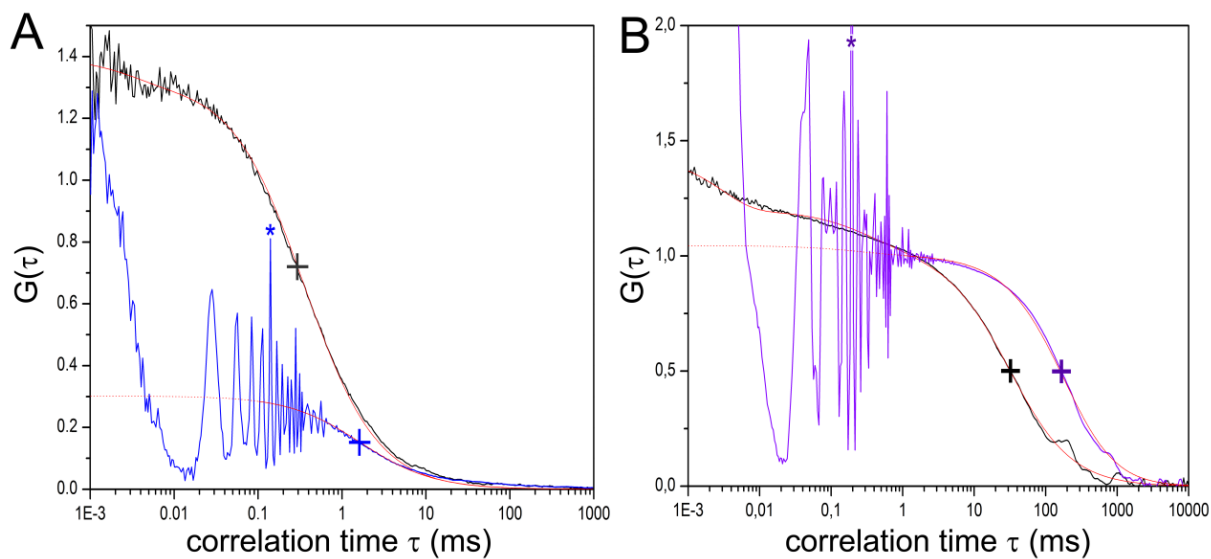

**Figure S9**. **A**, FCS of R6G in solution with 32-point knight tour pattern off (black curve) or with active pattern (blue curve). Blue asterisk denotes the correlation peak due to the 7 kHz repetition frequency of the pattern. Crosses indicate the mean diffusion times after fitting (red curves). **B**, normalized FCS of Cy3B-labeled $F_oF_1$ in liposomes with the 1 µm shallow region of the ABELtrap chip with 32-point knight tour pattern turned off (black curve) or pattern and electrode feedback active (purple curve). Purple asterisk denotes the correlation peak due to the 5 kHz pattern repetition frequency. Crosses indicate the mean diffusion times according to the fitting (red curves).

The corresponding fluorescence intensity time traces of these proteoliposomes recorded in the shallow trapping region are shown in Figure S10. Keeping the laser focus position fixed, freely diffusing single $F_oF_1$-ATP synthases yielded photon bursts with maximum peak counts around 150 counts per ms, with the expected strong variations from burst to burst (Figure S10 A, black trace). Turning the knight tour focus pattern on resulted in an estimated 5-fold drop of the peak intensities to around 30 counts per ms, with a few proteoliposomes exhibiting higher peak photon counts (due to aggregates of liposomes or reconstitution of more than one enzyme in a liposome). Obviously, the intensity drop was caused by expanding the excitation volume which reduced the mean excitation power per area, but also increased the observation time or diffusion time respectively (blue trace, Figure S10 B). A strong variability of the fluorescence peak intensities due to the arbitrary diffusive motion of the proteoliposomes through the focus remained for the expanded excitation volume.

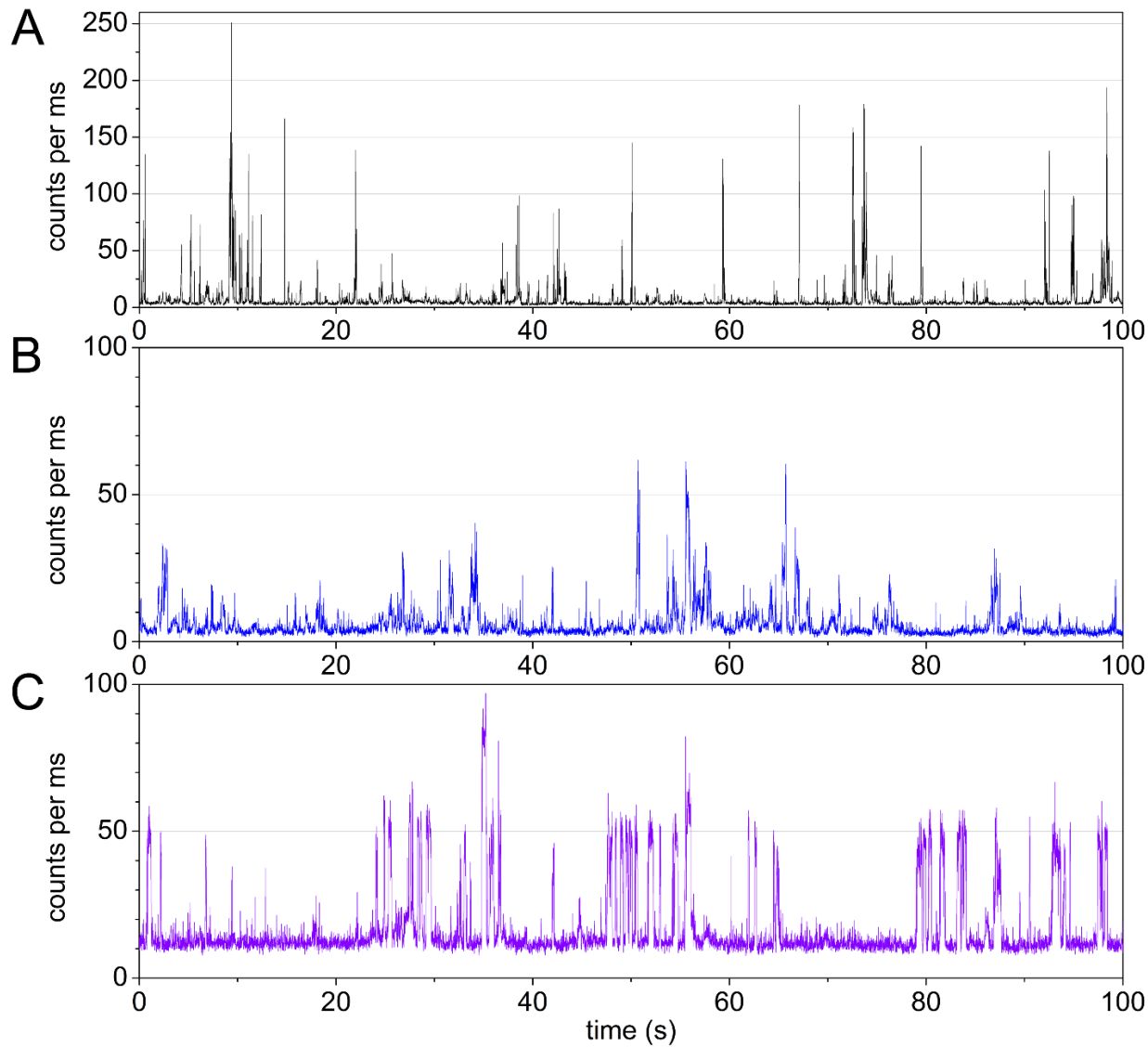

**Figure S10**. Fluorescence intensity time traces of FRET-labeled $F_oF_1$-ATP synthases in liposomes with 10 ms time binning, 532 nm excitation with 40 μW, intensities of FRET donor channel only. **A**, time trace with fixed laser focus position. **B**, time trace with 32-point knight tour focus pattern active. **C**, time trace with knight tour pattern and electrode feedback active, i.e., during ABEL trapping.

After starting the electrode feedback the duration and shape of the photon bursts changed (purple trace, Figure 10 C). The intensity variation within the burst and across bursts was much smaller, and the mean intensity level (~50 counts per ms) was higher than the average for freely diffusing proteoliposomes. However, also the background signal increased to ~10 counts per ms, resulting in a 4:1 signal-to-background ratio (note that the intensities were recorded in the FRET donor channel only in Figure S10).

**2.5. Photophysics of TMR, Cy3B and Atto R6G as FRET donor.** To select an optimal FRET donor fluorophore for 532 nm excitation with our ABEL trap setup reconstituted $F_oF_1$-ATP synthases were labeled at residue εH56C with either tetramethylrhodamine (TMR), Cy3B, or Atto Rho6G (or Atto R6G, respectively). First, molecular brightness was estimated using spectral data for the fluorophores as provided by FPbase (open source website *www.FPbase.org/spectra/*). For the previously used fluorophore TMR for smFRET with $F_oF_1$-ATP synthase [1, 3], the molar extinction coefficient at 532 nm was $\varepsilon_{(TMR,532)} = 53,000$ M$^{-1}$cm$^{-1}$ (i.e., 53% of the maximum $\varepsilon_{(TMR,552)} = 100,000$ M$^{-1}$cm$^{-1}$ at 552 nm), the fluorescence quantum yield was $\phi=0.34$ and a relative spectral detection efficiency limited by the optical filter 582/75 BrightLine HC was 0.765. For Cy3B, the molar extinction coefficient at 532 nm was $\varepsilon_{(Cy3B,532)} = 78,000$ M$^{-1}$cm$^{-1}$, $\phi=0.67$ and the spectral detection efficiency 0.692. For Atto R6G, the molar extinction coefficient at 532 nm was $\varepsilon_{(Atto\,R6G,532)} = 115,000$ M$^{-1}$cm$^{-1}$, $\phi=0.9$ and the spectral detection efficiency 0.757. For comparison, the molecular brightness of Cy3B was calculated as the product of $\varepsilon_{(532)} * \phi *$ [spectral detection efficiency] and normalized to 100%. Accordingly the expected relative molecular brightness for TMR was 38%, and for Atto R6G was 216%.

The experimental data in Figure S11 indicated that TMR attached to $F_oF_1$-ATP synthase exhibited a mean brightness of 14 counts per ms, Cy3B a mean brightness of 34 counts per ms, and Atto R6G a mean brightness of 41 counts per ms. The relative brightness of TMR was 41% with respect to Cy3B, or 121% for Atto R6G with respect to Cy3B. A substantial asymmetry of the brightness distribution was noticed for Atto R6G attached to $F_oF_1$-ATP synthase (Figure S11 H).

Despite the brightness differences of the three fluorophores, proteoliposomes were trapped with comparable probabilities. The burst duration distributions for TMR- or Atto R6G-labeled $F_oF_1$-ATP synthases were almost identical. Fitting the distributions with an exponential decay function yielded average trapping times of 350 ms for TMR and 370 ms for Atto R6G (1/e value), but Cy3B showed a significantly longer average trapping time of 750 ms. Therefore, Cy3B was chosen as the FRET donor for smFRET measurements in the ABEL trap.

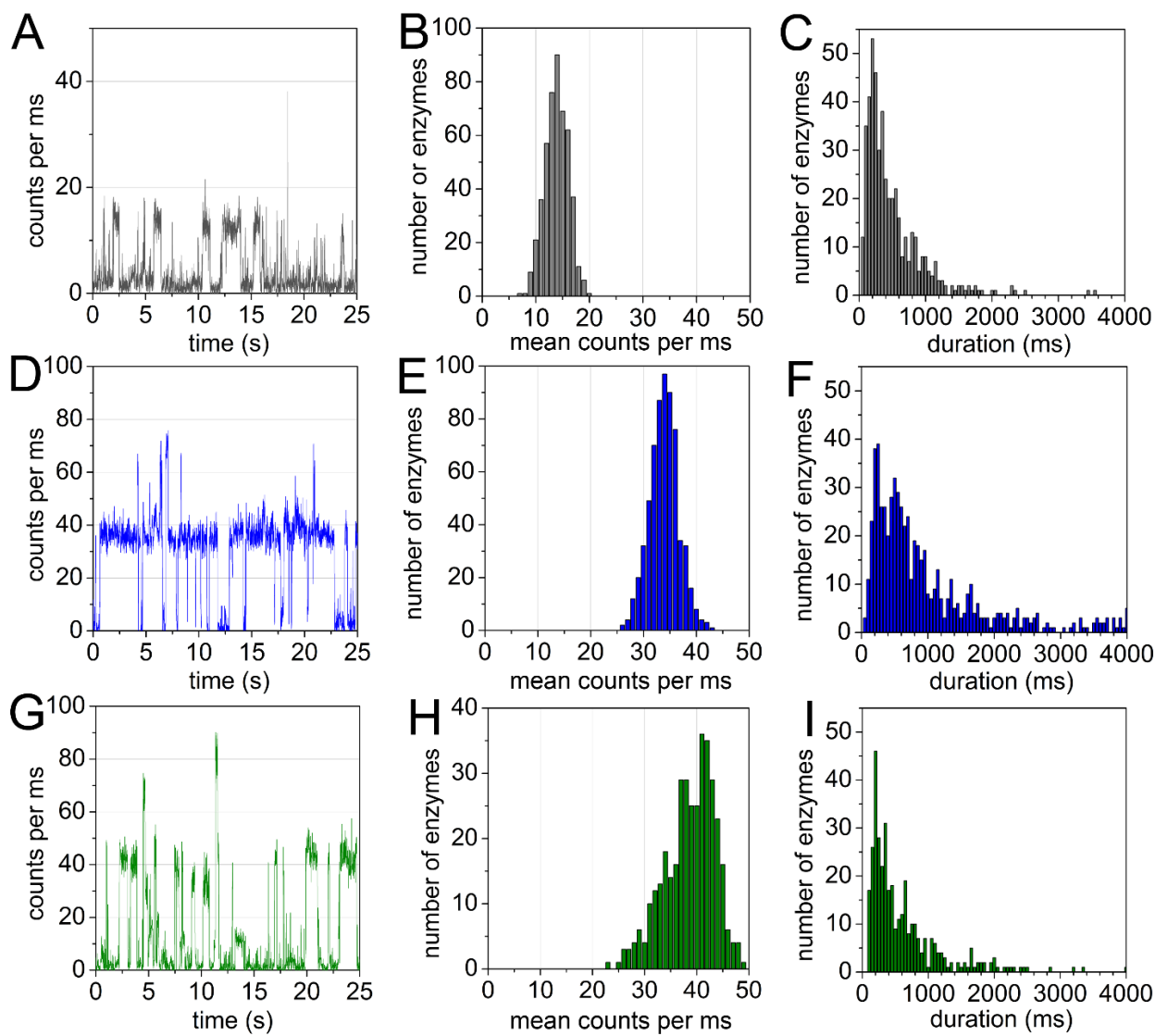

**Figure S11**. Comparison of TMR (grey), Cy3B (blue) and Atto R6G (green) attached to $F_oF_1$-ATP synthase in the ABEL trap. **A, D, G**, fluorescence intensity time traces with background subtracted, using 10 ms binning, recorded in the FRET donor channel. **B, E, H**, mean brightness distribution of ABEL trapped proteoliposomes. **C, F, I**, burst duration distribution of the ABEL trapped proteoliposomes.

Note that the mean count rates given in Figure S11 were smaller because only photons detected in the FRET donor channel were recorded. Therefore, using the total mean count rates in both FRET detection channel would have increased the mean count rates by 9% for TMR, by 16.5% for Cy3B and by 5.4% for Atto R6G.

## 3. Adding the antioxidant "*trolox*" as a triplet quencher

Photobleaching of the FRET acceptor Alexa Fluor 647 was observed frequently for FRET-labeled $F_oF_1$-ATP synthases in the ABEL trap, affecting the high FRET state more severely than the medium FRET state as noticeable by fewer long-lasting high FRET photon bursts.

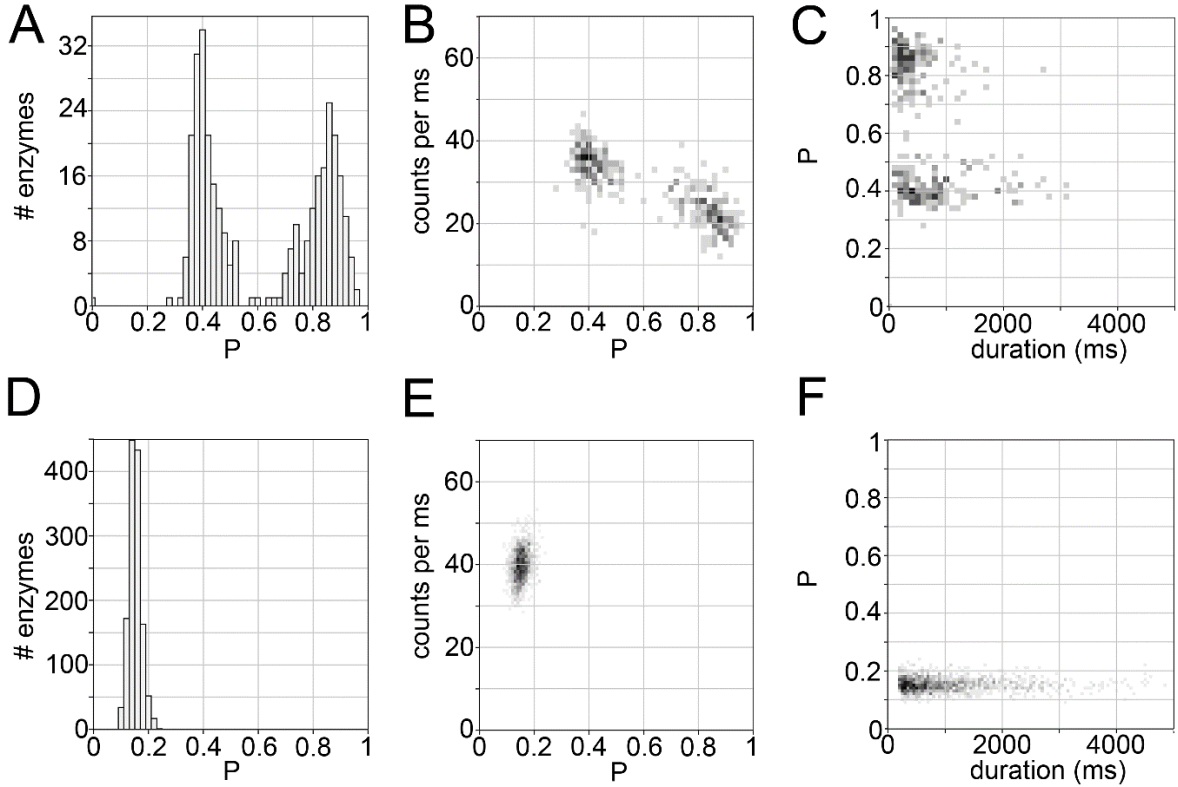

**Figure S12**. Proximity factor P distributions, related brightness and burst durations of proteoliposomes in the presence of 1 mM AMPNPN and 2 mM *trolox*. **A, B, C**, FRET-labeled $F_oF_1$-ATP synthases with Cy3B attached to εH56C and Alexa Fluor 647 to a-CT. **D, E, F,** "donor only"-labeled $F_oF_1$-ATP synthases.

To extend the trapping duration and to reduce photobleaching of Alexa Fluor 647, the addition of 2 mM *trolox* to the proteoliposomes in the presence of 1 mM AMPPNP (or 1 mM ATP, respectively) was evaluated. As shown in Figure S12 in the presence of 1 mM AMPPNP, the two FRET subpopulations exhibited almost the same proximity factor distributions and P-related brightness as in the absence of *trolox*. The medium FRET state subpopulation lasted slightly longer than the high FRET state subpopulation. Neither the mean brightness nor the trapping duration of "donor-only" enzymes with P ~ 0.15 did benefit from the addition of *trolox* so that adding this triplet quencher was not continued for the subsequent experiments on the ATP-dependent catalytic rates of single $F_oF_1$-ATP synthases.

## 4. Photon bursts at ATP concentrations of 100 µM, 40 µM, 20 µM, 5 µM

**4.1. Photon bursts at 100 µM ATP.** Two examples of FRET-labeled $F_oF_1$-ATP synthases exhibiting ATP hydrolysis-driven ε-subunit rotation in the presence of 100 µM ATP are shown in Figure S13 A, B. Throughout the photon bursts, FRET donor (green trace) and FRET acceptor (red trace) intensities changed in an anticorrelated pattern. The mean catalytic rate was calculated by counting the number of proximity factor fluctuations (i.e., a pair of one medium FRET level at P ~ 0.4 plus one high FRET level at P > 0.8, blue trace in the upper subpanel) during the duration of FRET fluctuations within the photon burst. The $EF_oF_1$ in Figure 13 A had an average turnover of 146 ATP·s$^{-1}$ as calculated from 74 full rotations in 1525 ms. The enzyme was recorded 22.7 min after start of the measurement (i.e., ~25 min after ATP addition). Apparently this ABEL-trapped enzyme did not operate at constant speed. Three phases could be distinguished: (1) at the beginning the turnover was slower with 82 ATP·s$^{-1}$ for 220 ms (6 full rotations), followed by (2) a fast catalytic phase with 185 ATP·s$^{-1}$ for 975 ms (60 full rotations), and (3) a slow phase for the remaining 330 ms with 73 ATP·s$^{-1}$ (8 full rotations). The $EF_oF_1$ in Figure 13 B was recorded 23.1 min after starting the ABEL trap measurement and exhibited an average catalytic rate of 40 ATP·s$^{-1}$ for 1365 ms. However, for the first 750 ms period the rate was only 12 ATP·s$^{-1}$, followed by a rate of 72 ATP·s$^{-1}$ before the FRET acceptor photobleached.

**4.2. Photon bursts at 40 µM ATP.** Two examples of catalytically active FRET-labeled $F_oF_1$-ATP synthases in the presence of 40 µM ATP are shown in Figure S13 C, D. The enzyme in Figure 13 C was recorded 22.7 min after starting the ABEL trap measurement and exhibited an average catalytic rate of 138 ATP·s$^{-1}$. After a 1109 ms period of fast FRET fluctuations, this $F_oF_1$-ATP synthase stopped turnover and remained in the medium FRET state (P ~ 0.4) for 260 ms before the FRET acceptor photobleached and the enzyme was lost from the ABEL trap. The second example in Figure 13 D exhibited an average catalytic rate of 164 ATP·s$^{-1}$ and was recorded 6.2 min after starting the ABEL trap measurement. During a first 230 ms period the rate was 225 ATP·s$^{-1}$, followed by a slower rate of 134 ATP·s$^{-1}$.

**4.3. Photon bursts at 20 µM ATP.** In Figure S13 E, two examples of catalytically active FRET-labeled $F_oF_1$-ATP synthases are shown back-to-back in the presence of 20 µM ATP. The first enzyme in the time trace exhibited an average catalytic rate of 29 ATP·s$^{-1}$ during an 833 ms period.

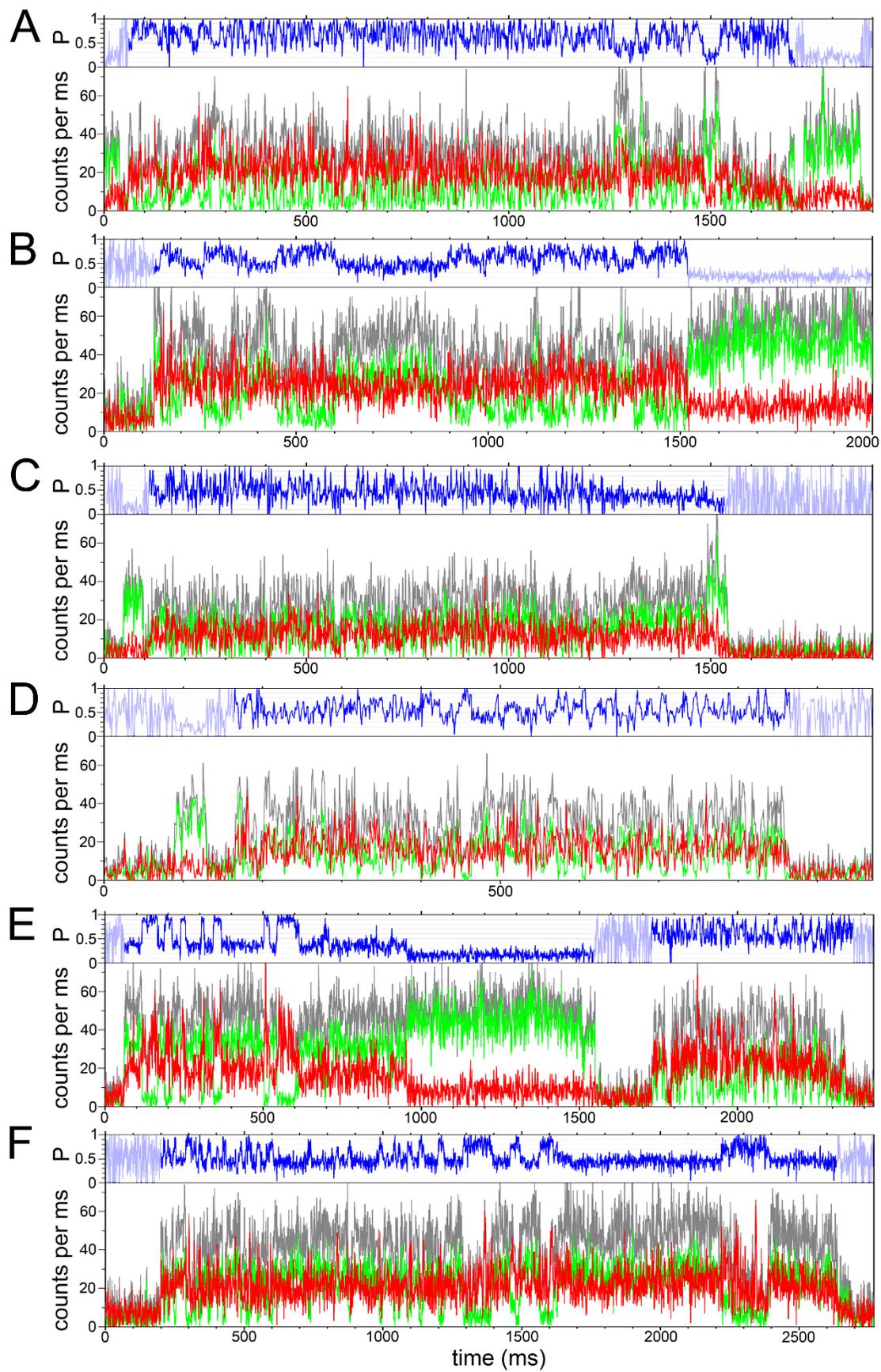

S-20

**Figure S13. A, B,** photon bursts of FRET-labeled $F_oF_1$-ATP synthases held in the ABEL trap with 100 µM ATP. **C, D,** photon bursts with 40 µM ATP. **E, F,** photon bursts with 20 µM ATP. FRET donor intensity as green traces, FRET acceptor as red traces, summed intensities as grey traces, and proximity factor P in blue traces. 1 ms time binning.

The two well-defined FRET level at P ~ 0.4 and P ~ 0.9 alternated with fast transitions (within 1 to 3 ms). After FRET acceptor photobleaching the "donor only"-labeled enzyme remained trapped for another 600 ms. The subsequent photon burst showed a $EF_oF_1$ comprising 14 rotations in 538 ms, i.e. yielding an average catalytic rate of 78 ATP·s$^{-1}$. Both enzymes were observed 22 min after starting the ABEL traps recording. In Figure S13 F, the $F_oF_1$-ATP synthase was recorded after 1.7 min and was trapped for 2420 ms. The enzyme showed an average turnover of 29 ATP·s$^{-1}$. However, at the beginning the catalytic rate was faster with 97 ATP·s$^{-1}$ or 13 full rotations in 404 ms, respectively. Then the catalytic rate dropped to 33 ATP·s$^{-1}$ for 1080 ms, and slowed down even further. In the next 955 ms only three FRET level were found i.e. the enzyme almost stopped catalysis.

**4.4. Photon bursts at 5 µM ATP.** The lowest substrate concentration used in the experiments was 5 µM ATP. In Figure S14, four examples of active FRET-labeled $F_oF_1$-ATP synthases held in solution by the ABEL trap are shown. In Figure S14 A, the enzyme recorded after 16.8 min, hydrolyzed ATP at an average rate of 19 ATP·s$^{-1}$ during 950 ms. For the first ~ 500 ms of the photon burst the enzyme was found in the medium-to-low FRET state and was apparently inactive before starting ATP hydrolysis. In Figure S14 B, the enzyme had a turnover of 24 ATP·s$^{-1}$ during 737 ms and was recorded 11.4 min after starting the ABEL trap measurement.

In contrast to fast FRET transition times of $F_oF_1$-ATP synthases in the presence of 20 µM ATP shown in Figures S13 E and F, switching between the medium FRET level at P ~ 0.4 and P ~ 0.85 occurred over several tens of ms and in a "ramp-like" behavior. In contrast, the $EF_oF_1$ in Figure S14 C exhibited fast FRET fluctuations yielding an average rate of 102 ATP·s$^{-1}$ during 704 ms before FRET acceptor photobleaching. The fast catalysis of this enzyme was considered an outlier with respect to the mean ATP hydrolysis rate of 47 ATP·s$^{-1}$ calculated from enzymes analyzed at 5 µM ATP. Another example of an active FRET-labeled $F_oF_1$-ATP synthase is shown in Figure S14 D. The ATP hydrolysis rate was 16 ATP·s$^{-1}$ during 1844 ms before FRET acceptor photobleaching. FRET level transitions seemed to comprise not only P ~ 0.4 and P > 0.8 but also an additional FRET level at $0.6 < P < 0.7$ with a duration of 20 to 80 ms. Accordingly, an apparent FRET level sequence of $\rightarrow(P > 0.8)\rightarrow(P \sim 0.65)\rightarrow(P \sim 0.4)\rightarrow(P > 0.8)\rightarrow$ could be assigned.

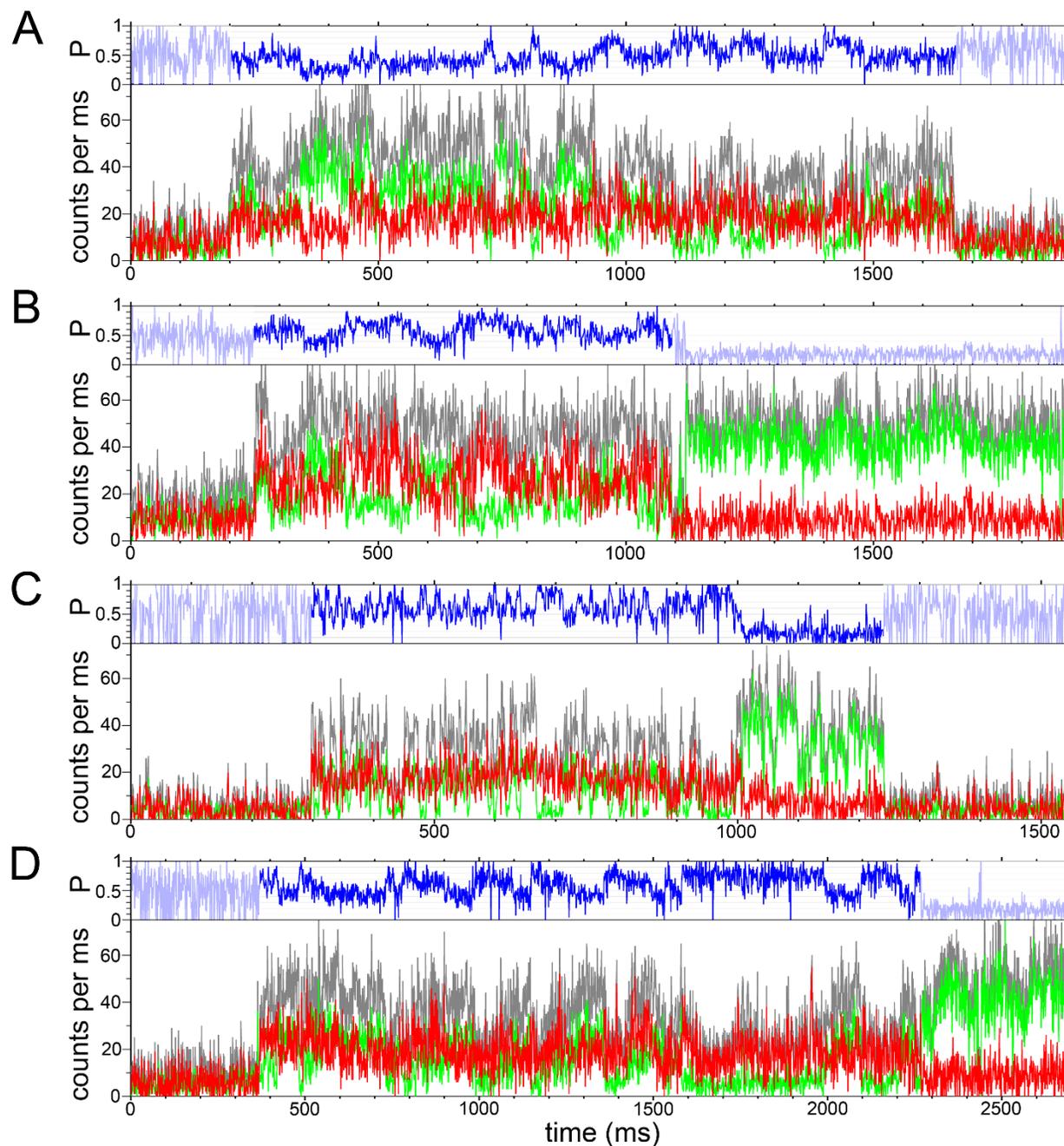

**Figure S14**. Four photon bursts of FRET-labeled $F_oF_1$-ATP synthases in the presence of 5 µM ATP. FRET donor intensity as green traces, FRET acceptor as red traces, summed intensities as grey traces, and proximity factor P as blue traces. 1 ms time binning.

**4.5. Duration of FRET fluctuations and related ATP hydrolysis rates**. The durations of active catalysis of proteoliposomes in the ABEL trap were investigated for different ATP concentrations. For comparison, the distribution of "donor only" labeled $F_oF_1$-ATP synthases, had been fitted by a monoexponential decay and a 1/e trapping duration of 750 ms had been calculated. In the presence of different ATP concentrations, the distributions of the FRET fluctuation periods for active enzymes appeared similar (Figure S15). Only a few enzymes were trapped for more than 1 second. Fitting the distribution in the presence of 1 mM ATP (Figure S15 A) yielded a 1/e duration of 350 ms for the mean catalytic activity, i.e. the distribution of the periods of catalytic activity were shorter than the trapping times of the "donor only" labeled $F_oF_1$-ATP synthases.

In the presence of 100 µM ATP and 50 µM CCCP, the durations of active catalysis were shorter than in the presence of the uncouplers valinomycin plus nigericin. (Figure S15 F, G). Because we noticed a slightly different fluorescence or impurity background for each PDMS-on-glass chip and the time-dependent bleaching of the background, these effects on the actual signal-to-background ratio might have caused a different trapping behavior for some chips and might explain the apparent variations of the histograms of the active turnover periods.

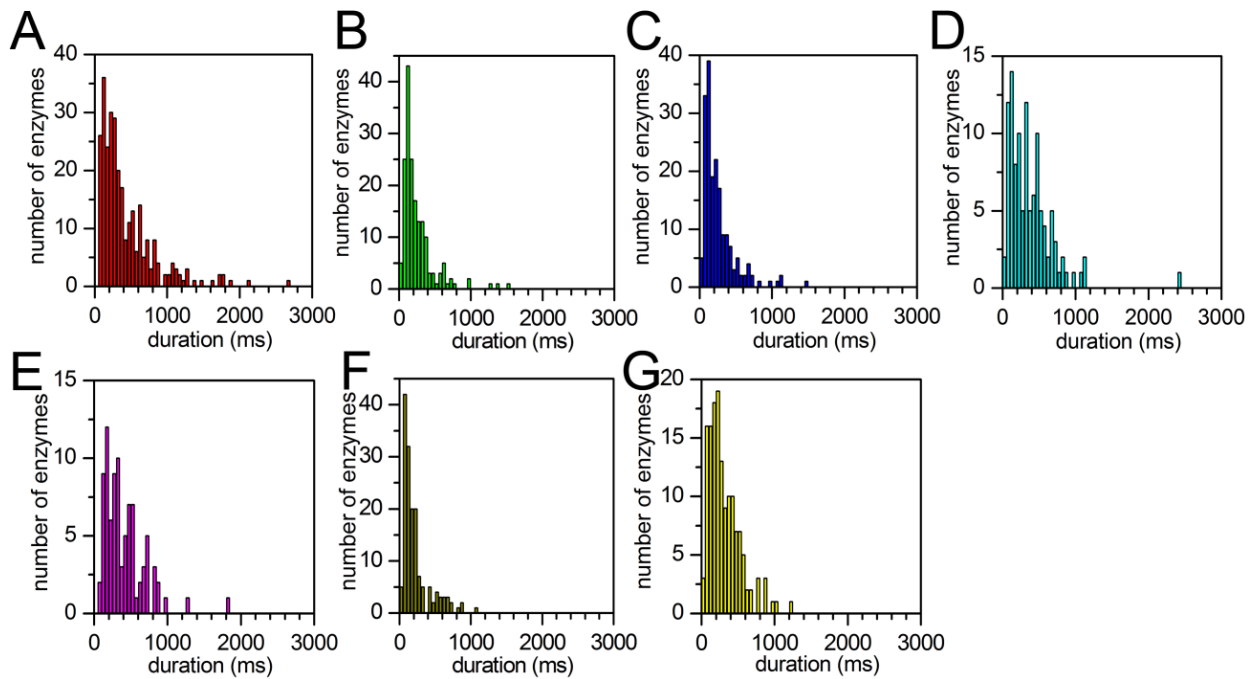

**Figure S15.** ATP-dependent distributions of FRET fluctuation periods from reconstituted $F_oF_1$-ATP synthases, in the presence of **A**, 1mM ATP, **B**, 100 µM ATP, **C**, 40 µM ATP, **D**, 20 µM ATP, **E**, 5 µM ATP, **F**, 100 µM ATP plus 50 mM CCCP, **G**, 100 µM ATP plus 1 µM valinomycin and 1 µM nigericin.

**4.6. Analyzing ATP hydrolysis rates by Michaelis-Menten kinetics**. The mean ATP hydrolysis rates obtained from measurements at five ATP concentrations (1 mM, 100 µM, 40 µM, 20 µM and 5 µM) were used to estimate the maximum turnover rate $V_{(max)}$ and the Michaelis-Menten constant $K_M$. Using the Lineweaver-Burk linearization [24] as shown in Figure S16, fitting (red line) resulted in $V_{(max)} = 147$ ATP·s$^{-1}$ from the intercept and $K_M = 11.7$ µM from the slope. In Figure S16, triangles are the limits defined by the 1$^{st}$ and 3$^{rd}$ quartile of the individual rate distributions from the box plot (see manuscript for details).

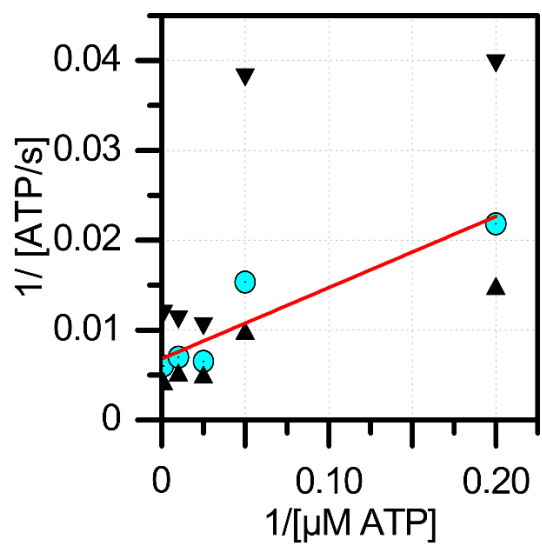

**Figure S16**. Lineweaver-Burk plot of ATP-dependent mean catalytic rates of FRET-labeled $F_oF_1$-ATP synthases held in solution by the ABEL trap (cyan circles; with upper and lower limits defined by the 1$^{st}$ and 3$^{rd}$ quartile of the individual rate distributions from the box plot, black triangles).

## 5. References for the Supporting Information